\documentclass[aps,pre,reprint,notitlepage,onecolumn, nofootinbib]{revtex4-1}

\usepackage{anysize}

\usepackage[utf8]{inputenc}

\usepackage{amsmath}
\usepackage{amssymb}
\usepackage{bm}
\usepackage{hyperref}
\usepackage{graphics,graphicx}

\usepackage{array}   
\newcolumntype{L}{>{$}l<{$}} 
\newcolumntype{C}{>{$}c<{$}} 
\newcolumntype{R}{>{$}r<{$}} 

\renewcommand{\i}{\alpha}
\renewcommand{\j}{\beta}

\marginsize{0.5in}{0.5in}{0.5in}{0.5in}

\begin{document}
\begin{abstract}
We provide the $21N$-moment multi-temperature collision coefficients for the Boltzmann collision operator using the Sonine-Hermite polynomial ansatz in the style of Zhdanov et al. First, we outline the general derivation method. Then, we provide the collision coefficients in the most general form in terms of the Chapman-Cowling integrals for any potential of interaction for the $21N$-moment approximation. Then, we provide the the collision coefficients for the specific case of the approximated Coulomb potential cross-section with Debye cutoff. These coefficients help in order to find easy implementation in current SOL/edge fluid packages which currently implement the $21N$-moment single-temperature coefficients. This provides the completion of a missing link in the current literature. 
\end{abstract}
 
\title{The $21N$-moment multi-temperature coefficients for Zhdanov closure}
\author{M.\,Raghunathan}
\author{Y.\,Marandet}
\affiliation{Aix-Marseille Univ., CNRS, PIIM, Marseille, France}
\author{H.\,Bufferand} 
\author{G.\,Ciraolo}
\author{Ph.\,Ghendrih}
\author{P.\,Tamain}
\affiliation{IRFM-CEA, F-13108 Saint-Paul-Lez-Durance, France}
\author{E.\,Serre}
\affiliation{Aix-Marseille Univ., CNRS, M2P2, Marseille, France}
\maketitle


\section{Introduction}

Understanding impurity dynamics in the scrape-off layer(SOL) and edge of tokamaks remains a key factor in improving the performance of current and next-step machines such a WEST, JET and ITER. For example, reducing the peak heat fluxes on plasma facing components to manageable levels relies on a large extent on impurity radiation, which in turn depends on the impurity density, flow, and temperature. Generally, the SOL/edge is modeled as a high-collisionality regime, and the plasma is generally simulated with fluid packages, normally coupled to kinetic neutrals, e.g. Soledge3x-EIRENE\cite{bufferand_2019}. Generally, the collision terms used in the fluid packages are found by averaging over the kinetic equation with different constants of motion. However, the procedure of such averaging typically leaves the system of equations obtained unclosed. Closing the system of equations is the procedure for obtaining these unclosed quantities, such as the collisional friction and thermal fornces, in terms of the plasmadynamical moments such as density, flow velocity and temperature. 

One such method of closure is called the Zhdanov closure\cite{zhdanov_effect_1962,yushmanov_diffusion_1980}, which involves using Grad's method\cite{grad_asymptotic_1963} with a Sonine-Hermite polynomial ansatz for the distribution functions and the moments. On averaging over the Boltzmann kinetic equation with the moments, and assuming low Knudsen number, one finds prescription for the friction and thermal forces in terms of the flow velocities and the temperature gradients. The advantage of this closure is that it does not depend on any trace density assumptions or low temperature difference assumptions, as opposed to earlier closures such as Braginskii's\cite{braginskii_transport_1965}. This is important for modelling impurity dynamics because impurities may be present as a significant fraction of the background plasma species, or, in case of deuterium-tritium plasmas, tritium, being a main ion species, cannot be modelled as trace over an existing deuterium-electron plasma. The Zhdanov closure is already implemented, though only using single-temperature collision coefficients at the plasma common temperature, in SOL/edge simulators such as Soledge3x-EIRENE\cite{bufferand_near_2013,bufferand_2019,bufferand_implementation_2021}, SOLPS\cite{rozhansky_momentum_2015,sytova_impact_2018,makarov_equations_2021}, B2-EIRENE\cite{fichtmuller_multi-species_1998}, and EDGE2D\cite{bergmann_implementation_1996}. 

Generally, the depth of the closure is represented by the number of moments, i.e. $5N$-moment, $13N$-moment, $21N$-moment or $29N$-moment, where $N$ represents the number of species. In Refs.\,\cite{raghunathan_generalized_2021} and \cite{raghunathan_multi-temperature_2021}, we found that generally, the multi-temperature $21N$-moment scheme exhibits values of transport coefficients and friction/thermal forces about 20-40\% different from the single-temperature scheme, and furthermore, is adequate for obtaining converged values of friction and thermal forces for fusion-relevant scenarios. However, the collision coefficients for the $21N$-moment multi-temperature Zhdanov closure scheme have not been provided explicitly in any literature. Thus, in this article, we provide the $21N$-moment multi-temperature collisional coefficients derived from the Boltzmann collision operator to fill the existing gap. We hope that the availability of these coefficients will lead to a straightforward implementation in current SOL/edge fluid packages.

The article is organized a follows. Sections \ref{sec:boltzmann} and \ref{sec:derivation_operator} give an overview of the moment averaging process over the collision operator using the Sonine-Hermite polynomial ansatz, describes all the relevant terminology, and derives expressions for the collision coefficients. In Section \ref{sec:multi-temp_coeffs}, we provide the values of the multi-temperature collision coefficients for the $21N$-moment scheme in the most general form in terms of Chapman-Cowling integrals, which are applicable to any potential of interaction. In Section \ref{sec:multi-temp_coeffs_coulomb}, we provide the values of the $21N$-moment collision coefficients for the Coulomb potential with Debye cutoff approximation. With these values, one should be able to implement Zhdanov closure for the multi-temperature case. In Section \ref{sec:summary}, we summarize the article.

\section{The Boltzmann equation for multi-species plasma}
\label{sec:boltzmann} 

The Boltzmann equation for the distribution function for species $\alpha$, $f_\alpha$ in the frame of the peculiar velocity of species $\i$, $\mathbf{c_\alpha}=\mathbf{v_\alpha}-\mathbf{u}$, is given by
\begin{equation}
 \frac{d f_\alpha}{d t}+\mathbf{c_\alpha}.\nabla{f_\alpha}+\frac{1}{m_\alpha}\mathbf{F}^*_\alpha.\nabla_{c_\alpha}{f_\alpha} -c_{\alpha s}\frac{\partial f_\alpha}{\partial c_{\alpha_r}}\frac{\partial u_r}{\partial x_s}= \sum_{\beta} J_{\alpha\beta}, \label{boltzmannc}
\end{equation} 
where the common plasma flow velocity $\mathbf{u}$ is given by
\begin{equation}
 \rho \mathbf{u} = \sum_\alpha \rho_\alpha \mathbf{u_\alpha},\ \rho = \sum_\alpha \rho_\alpha,
\end{equation}
where $\rho$ represents the mass density.The $d/dt$ represents full time derivative given by $d/dt=\partial/\partial t+\mathbf{u}.\nabla$, and where the force term $\mathbf{F_\alpha}$ and $d\mathbf{u}/dt$ are combined to write the relative force in the moving frame given by $\mathbf{F^*_\alpha}=\mathbf{F_\alpha}-m_\alpha d\mathbf{u}/dt$.

The LHS is referred to as the ``free-streaming term'', and the RHS is the collisional contribution between species $\i$ and every other species of the system. The collisional RHS is represented by the Boltzmann collision integral,
\begin{equation}
 J_{\alpha\beta} = \iint (f^\prime_\alpha f^\prime_{1\beta}-f_\alpha f_{1\beta})g\sigma_{\alpha\beta}(g,\chi)d\Omega d\mathbf{c_{1\beta}}, 
\end{equation}
where $\alpha,1\beta$ refer to species of the two particles colliding, subscript $\prime$ refers to properties after the collision, $g$ is the relative velocity between the colliding particles, $\sigma_{\alpha\beta}$ is the collision cross section, and $\Omega$ is the solid angle in which the collision occurs. 



Now, for any quantity $\psi_\alpha$ depending purely on species $\alpha$, one can average over the RHS of Eq.~\ref{boltzmannc} which attains the following form
\begin{equation}
 R_\alpha=\sum_{\beta} \int \psi_\alpha J_{\alpha\beta}d\mathbf{v_\alpha}
 =\sum_{\beta}\iiint \psi_\alpha(f^\prime_\alpha f^\prime_{1\beta}-f_\alpha f_{1\beta})g\sigma_{\alpha\beta}(g,\chi)d\Omega d\mathbf{c_{\alpha}}d\mathbf{c_{1\beta}}.
\end{equation}
For elastic collisions, the moment averaged collision operator can be transformed into
\begin{equation}
 R_\alpha=\sum_{\beta}\iiint (\psi^\prime_\alpha-\psi_\alpha)f_\alpha f_{1\beta}g\sigma_{\alpha\beta}(g,\chi)d\Omega d\mathbf{c_{\alpha}}d\mathbf{c_{1\beta}},
 \label{eq:boltzmann2}
\end{equation}
since the distribution functions for any given species are purely a function of the species peculiar velocity.

We now describe the modification of Grad's method\cite{grad_asymptotic_1963} as used by Zhdanov\cite{zhdanov_transport_2002} in his previous papers. In the ansatz for the solution of the Boltzmann equation, it is assumed that the solution $f_\alpha$ is already near thermodynamic equilibrium for species $\alpha$, $f_\alpha^{(0)}$ as follows
\begin{align}
 f_\alpha^{(0)} &= n_\alpha \left( \frac{m_\alpha}{2\pi kT_\alpha}\right)^{3/2}\exp{\left(-\frac{m_\alpha c_\alpha^2}{2kT_\alpha}\right)}\nonumber\\
 &=n_\alpha\left( \frac{\gamma_\alpha}{2\pi} \right)^{3/2}\exp{\left( -\frac{\gamma_\alpha}{2}c_\alpha^2\right)},
 \label{eq:ansatz}
\end{align}
where $\gamma_\alpha={m_\alpha}/{kT_\alpha}$, $n_\i$ is the number density, and $T_\i$ the temperature of species $\i$. In order to solve the Boltzmann equation (\ref{boltzmannc}), Zhdanov and Yushmanov choose an ansatz of the form
\begin{equation}
f_\alpha = f_\alpha^{(0)}\sum_{m,n} 2^{2n}m_\alpha^{-2}\gamma_\alpha^{2n+m} \tau_{mn} b^{mn}_{\alpha r_1\ldots r_m}G^{mn}_{\alpha r_1\ldots r_m},\label{eq:ansatz}
\end{equation}
where
\begin{equation}
 G_\alpha^{mn}(\mathbf{c_\alpha},\gamma_\alpha) = (-1)^n n! m_\alpha\gamma_\alpha^{-(n+m/2)}S^n_{m+1/2}\left(\frac{\gamma_\alpha}{2}\mathbf{c}^2_\alpha\right)P^{(m)}(\gamma_\alpha^{1/2}\mathbf{c_\alpha}).
 \label{eq:sonine-hermite}
\end{equation}
Here, $S^n_{m+1/2}$ are the Sonine polynomials, given by,
\begin{equation}
 S^n_{m+1/2}\left(\frac{\gamma_\alpha}{2}\mathbf{c}^2_\alpha\right) = \sum_{p=0}^n \left(-\frac{\gamma_\alpha}{2}\mathbf{c}^2_\alpha\right)^p\frac{(m+n+1/2)!}{p!(n-p)!(m+p+1/2)!}.\nonumber
\end{equation}

Further, $P^{(m)}$ are the irreducible projection of the tensorial monomial $\mathbf{c}_\alpha^m=c_{\alpha r_1}\ldots c_{\alpha r_m}$, derived by the  following recurrence relation
\begin{equation}
 P^{(m+1)}(\gamma_\alpha^{1/2}\mathbf{c_\alpha}) = \gamma_\alpha^{1/2}\mathbf{c_\alpha} P^{(m)} -\gamma_\alpha^{1/2}\frac{c_\alpha^2}{2m+1}\frac{\partial P^{(m)}}{\partial \mathbf{c_\alpha}},\nonumber
\end{equation}
with $P^{(0)}=1$. 
The constant $\tau_{mn}$ arises as a result of internal contractions between $b_\alpha^{mn}$ and $G_\alpha^{mn}$, and is given by
\begin{equation}
 \tau_{mn}= \frac{(2m+1)!(m+n)!}{n!(m!)^2(2m+2n+1)!}. \nonumber
\end{equation}

The forms mentioned in Refs.~\cite{grad_asymptotic_1963} and \cite{zhdanov_transport_2002} are cosmetically different because of the choice to use full factorial representations of functions and because of summing over full indices rather than over half-indices, but they are equivalent. The coefficients $b^{mn}_\alpha$ are calculated as
\begin{equation}
 n_\alpha b^{mn}_\alpha = \int G_\alpha^{mn} f_\alpha d\mathbf{c}_\alpha.
\end{equation}
Some values of $G_\alpha^{mn}$ are as follows
\begin{align}
 G^{00}_\alpha &= m_\alpha,\ G^{10}_\alpha = m_\alpha \mathbf{c}_\alpha, \ G^{01}_\alpha=\frac{m_\alpha}{2}\left( c_\alpha^2-\frac{3}{\gamma_\alpha}\right),\nonumber\\
  G^{11}_\alpha &= \frac{m_\alpha}{2}\mathbf{c}_\alpha \left( c_\alpha^2-
 \frac{5}{\gamma_\alpha}\right),\ G^{20}_\alpha=m_\alpha\left(\mathbf{c}_\alpha\mathbf{c}_\alpha-\frac{1}{3}Uc_\alpha^2\right)\nonumber\\
 G^{12}_\alpha &=\frac{m_\alpha}{4}\mathbf{c}_\alpha(c_\alpha^4-14\gamma_\alpha^{-1}c_\alpha^2+35\gamma_\alpha^{-12}),\nonumber\\
 G^{21}_\alpha&=\frac{m_\alpha}{2}(c_\alpha^2-7\gamma_\alpha^{-1})\left(\mathbf{c}_\alpha\mathbf{c}_\alpha-\frac{1}{3}Uc_\alpha^2\right),\nonumber
\end{align}
and the corresponding $b^{mn}_\alpha$ are given by
\begin{align}
 n_\alpha b^{00}_\alpha &= \rho_\alpha,\ n_\alpha b^{10}_\alpha = \rho_\alpha \mathbf{w}_\alpha,\ n_\alpha b^{01}_\alpha = 0,\nonumber\\
 n_\alpha b^{11}_\alpha &= \mathbf{h}_\alpha,\ n_\alpha b^{20}_\alpha = \pi_\alpha,\nonumber\\
 n_\alpha b^{12}_\alpha &= \mathbf{r}_\alpha,\ n_\alpha b^{21}_\alpha = \sigma_\alpha.\nonumber
\end{align}
Here, $b^{00}_\alpha,\ b^{01}_\alpha$ and $b^{10}_\alpha$ represent the intuitive hydrodynamical moments density, diffusion velocity and temperature $\rho_\alpha$, $\mathbf{w}_\alpha=\mathbf{u}_\i-\mathbf{u}$, and $T_\alpha$. the higher moments $b^{11}_\alpha$ and $b^{20}_\alpha$ represent the thermodynamically privileged moments (as per Balescu's nomenclature\cite{balescu_transport_1988}), the heat flux $\mathbf{h}_\alpha$ and the divergence free pressure-stress tensor $\pi_\alpha$, which are called ``privileged'' because they contribute to the entropy production. The higher-order moments $b^{12}_\alpha$ and $b^{21}_\alpha$ are non-privileged moments $\mathbf{r}_\alpha$ and $\sigma_\alpha$, which do not have a clear physical meaning, however which may contribute to the accuracy of moment equations in terms of representing the Boltzmann equation. As one can notice, these are all moments of ranks less than 2. Moments of rank-0 are scalar, like density $\rho_\i$ and temperature $T_\i$, are constitute $N$ variables each. Moments rank-1 are vectorial moments, like momentum $m_\i\mathbf{w}_\i$ and heat-flux $\mathbf{h}_\i$, and contribute $3N$ variables each. Moments of rank-2 are tensorial in nature, like the stress-tensor $\pi_\i$ and $\sigma_\i$, and contribute $5N$ variables each (and not $9N$, since they are symmetric and traceless). In principle, one can construct a $5N$-moment system with just the hydrodynamical moments, a $13N$-system with including the thermodynamically privileged moments, and a $21N$-system including $\mathbf{r}_\i$ and $\sigma_\i$.

\section{Derivation of the right hand side of the Boltzmann equation}
\label{sec:derivation_operator}

In the Boltzmann collision integral $J_{\alpha\beta}$, it is possible to choose a distribution function which takes the form
\begin{equation}
 f_\alpha = f^{(0)}_\alpha (1+\Phi_\alpha), \nonumber
\end{equation}
which essentially represents the ansatz as a perturbed Maxwellian. The linearized moment-averaged collision operator Eq.\,(\ref{eq:boltzmann2}) can then be written as
\begin{equation}
 R_{\alpha\beta}=\int \psi_\alpha J_{\alpha\beta}d\mathbf{c_{\alpha}}\iiint f^{(0)}_\alpha f^{(0)}_\beta(\psi^\prime_\alpha-\psi_\alpha)(1+\Phi_\alpha+\Phi_\beta)g\sigma_{\alpha\beta}(g,\chi)d\Omega d\mathbf{c_{\alpha}}d\mathbf{c_{1\beta}}, \nonumber
\end{equation}
 
This allows us to decompose the moment-average into sums of smaller terms, which is useful analytically. This also is similar to the properties exhibited by some other linearized operators such as the linearized Landau operator\cite{balescu_transport_1988,helander_collisional_2005}. On substituting the Sonine-Hermite polynomial ansatz from Eq.\,(\ref{eq:ansatz}) for the distribution functions $f$, and set the moment $\psi=G^{mn}_{\i}$ from Eq.\,(\ref{eq:sonine-hermite}), we obtain
\begin{equation}
 R^{mnkl}_{\alpha\beta} = \iiint f^{(0)}_\alpha f^{(0)}_\beta \{G^{mn}_\alpha(\mathbf{c}_\alpha^\prime)-G^{mn}_\alpha(\mathbf{c}_\alpha)\}\{1+
 2^{2l}\gamma_\alpha^{2l+k}m_\alpha^{-2}\tau_{kl}G_\alpha^{kl}(\mathbf{c}_\alpha)b^{kl}_\alpha\\
 +2^{2l}\gamma_\beta^{2l+k}m_\beta^{-2}\tau_{kl}G_\beta^{kl}(\mathbf{c}_\beta)b^{kl}_\beta
 \}g\sigma_{\alpha\beta}(g,\chi)d\Omega d\mathbf{c_{\alpha}}d\mathbf{c_{1\beta}}, \nonumber
\end{equation} 
where $R^{mnkl}_{\alpha\beta}$ represents the part of $R^{mn}_{\alpha\beta}$ averaging over the $kl$ term of the ansatz Eq.\,(\ref{eq:ansatz}).  Noting that $G^{kl}b^{kl}$ is an inner product, we now substitute the definition of $G^{mn}$, and use the following integral identity\cite{ji_exact_2006,weinert_spherical_1980}
\begin{equation}
 \int P^{(m)}(P^{(k)}:W)G(v)d\mathbf{v}
 =\frac{W}{2m+1}\delta_{km}\int P^{(m)}:P^{(m)}G(v)d\mathbf{v}, \label{eq:pmidentity}
\end{equation}
where $W$ is symmetric and traceless tensor of rank $k$ not a function of $\mathbf{v}$. Furthermore, we define a ``bracket'' integrals of the following form
\begin{equation}
 n_\alpha n_\beta [F,G]=\iiint  f^{(0)}_\alpha f^{(0)}_\beta G(F^\prime-F)g\sigma_{\alpha\beta}(g,\chi)d\Omega d\mathbf{c_{\alpha}}d\mathbf{c_{1\beta}},
\end{equation}
through which we can contract over index $k$ in $R^{mnkl}_{\alpha\beta}$ and write it as $R^{mnl}_{\alpha\beta}$, such that 
\begin{equation}
 R_{\alpha\beta}^{mnl}=(1-\delta_{m0}\delta_{l0})(A_{\alpha\beta}^{mnl}b^{ml}_\alpha+B_{\alpha\beta}^{mnl}b^{ml}_\beta)+\delta_{m0}\delta_{l0}C_{\alpha\beta}^{mnl},
 \label{eq:collision_second_form}
\end{equation}
where $A_{\alpha\beta}^{mnl}$, $B_{\alpha\beta}^{mnl}$ and $C_{\alpha\beta}^{mnl}$ are given by
\begin{align}
 A_{\alpha\beta}^{mnl}&=  Q_{\alpha\beta}^{mnl}  
 \gamma_\alpha^{l-n}\left[S^n_{m+1/2}\left(W_\alpha^2\right)P^{(m)}(\mathbf{W}_\alpha),S^l_{m+1/2}(W_\alpha^2)P^{(m)}(\mathbf{W}_\alpha)\right]\nonumber\\
B_{\alpha\beta}^{mnl}&=Q_{\alpha\beta}^{mnl}\frac{\gamma_\beta^{l+m/2}}{\gamma_\alpha^{n+m/2}}\frac{m_\alpha}{m_\beta}\left[S^n_{m+1/2}(W_\alpha^2)P^{(m)}(\mathbf{W}_\alpha),S^l_{m+1/2}(W_\beta^2)P^{(m)}(\mathbf{W}_\beta)\right]\nonumber\\
 C^{mnl}_{\alpha\beta} &= (-1)^n n! \gamma_\alpha^{-(n+m/2)}m_\alpha n_\i n_\j\left[S^n_{1/2}(W_\alpha^2),1\right],
 \label{eq:AmnBmn1}
\end{align}
where
\begin{equation}
 Q_{\alpha\beta}^{mnl}=(-1)^{n+l}2^{2l+m} \frac{(2m)!(m+l)!n!}{(m!)^2 (2m+2l+1)!}n_\alpha n_\beta.
 \label{eq:qmnl}
\end{equation}
This moment-averaged collision operator is valid for any difference of masses or temperatures of the colliding species. The general expressions for the collision coefficients $A^{mnl}_{\i\j}$, $B^{mnl}_{\i\j}$, and $C^{mnl}_{\i\j}$ take the following general forms 
 \begin{equation}
 A^{mnl}_{\i\j},\ B^{mnl}_{\i\j},\ C^{mnl}_{\i\j}
 \sim\sum_{rl} \mathcal{A}^{pqrl,m}_{\i\i}\Omega_{\i\j}^{lr},\ \sum_{rl} \mathcal{A}^{pqrl,m}_{\i\j}\Omega_{\i\j}^{lr},\ \sum_{rl} \mathcal{A}^{pqrl,m}\Omega_{\i\j}^{lr}\nonumber
 \end{equation}
respectively. The $\mathcal{A}$-coefficients are expressed as functions of mass and temperature ratios of the species $\i$ and $\j$. The terms $\Omega_{\i\j}^{lr}$ are the effective cross-section moment integrals of Chapman and Cowling (henceforth referred to as the ``Chapman-Cowling integrals'') which are dependent on the potential of interaction between species $\i$ and $\j$. 
The exact derivations of the generalized coefficients $A^{pqrl,m}_{\i\j}$ and $A^{pqrl,m}_{\i\i}$ for different bracket integrals up to rank-2, with all steps supplied for verification purposes, are provided in Ref. \onlinecite{raghunathan_generalized_2021}. 

{ The Chapman-Cowling integral is written in the following form for our case
\begin{align}
 \Omega_{\i\j}^{lr} =& \left(\frac{2\pi}{\gamma_{\i\j}}\right)^{1/2}\int_0^\infty \exp(-\zeta^2)   \zeta^{2r+3} \phi^{(l)}_{\i\j} d\zeta,\\
 \phi^{(l)}_{\i\j}=&\int^\infty_0 (1-\cos^l{\chi}) \sigma_{\alpha\beta}(g,\chi)\sin{\chi}d\chi,
\end{align}
where $\zeta=(\gamma_{\i\j}/2)^{1/2}g$.

\section{Multi-temperature collision coefficients in terms of the Chapman-Cowling integrals}
\label{sec:multi-temp_coeffs}

In this section, the following notation is used in addition to the ones introduced in the previous sections,
\begin{equation}
 \eta_{\i\j}=\frac{m_\i}{m_\j},\ \mathrm{and}\ \theta_{\i\j}=\frac{T_\i}{T_\j},
\end{equation}
for the simplicity of representing the collision coefficients. The rank-1 coefficients up to $p,q\leq1$ and rank-2 coefficients for $p,q=0$ are exactly the same as the ones provided by Zhdanov et al in Chapter 4 of Ref.\,\cite{zhdanov_transport_2002}, represented slightly differently. For the case of equal temperatures, i.e.\,$\theta_{\i\j}=1$, they are exactly the same as the transport coefficients provided in Chapter 8 of Ref.\,\cite{zhdanov_transport_2002} (with minor corrections which can be found in Appendix D of Ref.\,\cite{raghunathan_generalized_2021}). As mentioned earlier, the coefficients in this section are valid for any potential of interaction, and we provide the specific ones for the Coulomb potential in the next section. 

\subsection{Value of $C^{010}_{\i\j}$ for $p,q=0$}

\begin{multline}
p=1,q=0:\   \mathcal{A}^{1011,0} \Omega_{\i\j}^{11}\\
\mathcal{A}^{1011,0}=-\frac{16 k \eta _{\alpha \beta } \left(\theta _{\alpha \beta }-1\right) n_{\alpha } n_{\beta } T_{\alpha }}{\left(\eta _{\alpha \beta }+1\right){}^2 \theta _{\alpha \beta }}
\end{multline}

\subsection{Values of $B^{1pq}_{\i\j}$ for $p,q\leq2$}

\begin{multline}
 p=0,q=0:\   \mathcal{A}^{0011,1}_{\i\j} \Omega_{\i\j}^{11}\\
 \mathcal{A}^{0011,1}_{\i\j}=\frac{16 \eta _{\alpha \beta } n_{\alpha } n_{\beta }}{3 \eta _{\alpha \beta }+3}
\end{multline}

\begin{multline}
p=0,q=1:\   \mathcal{A}^{0111,1}_{\i\j} \Omega_{\i\j}^{11}+\mathcal{A}^{0121,1}_{\i\j} \Omega_{\i\j}^{12}\\
 \mathcal{A}^{0111,1}_{\i\j}=  -\frac{16 \eta _{\alpha \beta }^2 m_{\beta } n_{\alpha } n_{\beta }}{3 k \left(\eta _{\alpha \beta }+1\right) T_{\beta } \left(\eta _{\alpha \beta }+\theta _{\alpha \beta }\right)}\\
\mathcal{A}^{0121,1}_{\i\j}=\frac{32 \eta _{\alpha \beta }^2 m_{\beta } n_{\alpha } n_{\beta }}{15 k \left(\eta _{\alpha \beta }+1\right) T_{\beta } \left(\eta _{\alpha \beta }+\theta _{\alpha \beta }\right)}
\end{multline}

\begin{multline}
p=1,q=0:\  \mathcal{A}^{1011,1}_{\i\j} \Omega_{\i\j}^{11}+\mathcal{A}^{1021,1}_{\i\j} \Omega_{\i\j}^{12}+\mathcal{A}^{1022,1}_{\i\j} \Omega_{\i\j}^{22}\\
\mathcal{A}^{1011,1}_{\i\j}=-\frac{40 k \eta _{\alpha \beta } n_{\alpha } n_{\beta } T_{\alpha } \left(\eta _{\alpha \beta } \left(3 \theta _{\alpha \beta }-2\right)+\theta _{\alpha \beta }\right)}{3 \left(\eta _{\alpha \beta }+1\right){}^2
   m_{\alpha } \left(\eta _{\alpha \beta }+\theta _{\alpha \beta }\right)}\\
\mathcal{A}^{1021,1}_{\i\j}=\frac{16 k \eta _{\alpha \beta } n_{\alpha } n_{\beta } T_{\alpha } \left(\eta _{\alpha \beta }^2 \left(3 \theta _{\alpha \beta }^2-6 \theta _{\alpha \beta }+4\right)+2 \eta _{\alpha \beta } \theta _{\alpha \beta
   }+\theta _{\alpha \beta }^2\right)}{3 \left(\eta _{\alpha \beta }+1\right){}^3 \theta _{\alpha \beta } m_{\alpha } \left(\eta _{\alpha \beta }+\theta _{\alpha \beta }\right)}\\
\mathcal{A}^{1022,1}_{\i\j}=\frac{32 k \eta _{\alpha \beta }^2 \left(\theta _{\alpha \beta }-1\right) n_{\alpha } n_{\beta } T_{\alpha }}{3 \left(\eta _{\alpha \beta }+1\right){}^3 \theta _{\alpha \beta } m_{\alpha }}
\end{multline}

\begin{multline}
p=1,q=1:\  \mathcal{A}^{1111,1}_{\i\j} \Omega_{\i\j}^{11}+\mathcal{A}^{1121,1}_{\i\j} \Omega_{\i\j}^{12}+\mathcal{A}^{1131,1}_{\i\j} \Omega_{\i\j}^{13}+\mathcal{A}^{1122,1}_{\i\j} \Omega_{\i\j}^{22}+\mathcal{A}^{1132,1}_{\i\j} \Omega_{\i\j}^{23}\\
\mathcal{A}^{1111,1}_{\i\j}=\frac{8 \eta _{\alpha \beta } \theta _{\alpha \beta } n_{\alpha } n_{\beta } \left(\eta _{\alpha \beta } \left(21 \theta _{\alpha \beta }-10\right)+11 \theta _{\alpha \beta }\right)}{3 \left(\eta _{\alpha \beta
   }+1\right){}^2 \left(\eta _{\alpha \beta }+\theta _{\alpha \beta }\right){}^2}\\
\mathcal{A}^{1121,1}_{\i\j}=-\frac{32 \eta _{\alpha \beta } n_{\alpha } n_{\beta } \left(\eta _{\alpha \beta }^2 \left(21 \theta _{\alpha \beta }^2-26 \theta _{\alpha \beta }+10\right)+2 \eta _{\alpha \beta } \theta _{\alpha \beta } \left(8
   \theta _{\alpha \beta }-3\right)+5 \theta _{\alpha \beta }^2\right)}{15 \left(\eta _{\alpha \beta }+1\right){}^3 \left(\eta _{\alpha \beta }+\theta _{\alpha \beta }\right){}^2}\\
\mathcal{A}^{1131,1}_{\i\j}=\frac{32 \eta _{\alpha \beta } n_{\alpha } n_{\beta } \left(\eta _{\alpha \beta }^2 \left(3 \theta _{\alpha \beta }^2-6 \theta _{\alpha \beta }+4\right)+2 \eta _{\alpha \beta } \theta _{\alpha \beta }+\theta
   _{\alpha \beta }^2\right)}{15 \left(\eta _{\alpha \beta }+1\right){}^3 \left(\eta _{\alpha \beta }+\theta _{\alpha \beta }\right){}^2}\\
\mathcal{A}^{1122,1}_{\i\j}=-\frac{32 \eta _{\alpha \beta } n_{\alpha } n_{\beta } \left(\eta _{\alpha \beta } \left(7 \theta _{\alpha \beta }-5\right)+2 \theta _{\alpha \beta }\right)}{15 \left(\eta _{\alpha \beta }+1\right){}^3 \left(\eta
   _{\alpha \beta }+\theta _{\alpha \beta }\right)}\\
\mathcal{A}^{1132,1}_{\i\j}=\frac{64 \eta _{\alpha \beta }^2 \left(\theta _{\alpha \beta }-1\right) n_{\alpha } n_{\beta }}{15 \left(\eta _{\alpha \beta }+1\right){}^3 \left(\eta _{\alpha \beta }+\theta _{\alpha \beta }\right)}
\end{multline}

\begin{multline}
p=0,q=2:\    \mathcal{A}^{0211,1}_{\i\j} \Omega_{\i\j}^{11}+\mathcal{A}^{0221,1}_{\i\j} \Omega_{\i\j}^{12}+\mathcal{A}^{0231,1}_{\i\j} \Omega_{\i\j}^{13}\\
\mathcal{A}^{0211,1}_{\i\j}=\frac{8 \eta _{\alpha \beta }^3 m_{\beta }^2 n_{\alpha } n_{\beta }}{3 k^2 \left(\eta _{\alpha \beta }+1\right) T_{\beta }^2 \left(\eta _{\alpha \beta }+\theta _{\alpha \beta }\right){}^2}\\
\mathcal{A}^{0221,1}_{\i\j}=-\frac{32 \eta _{\alpha \beta }^3 m_{\beta }^2 n_{\alpha } n_{\beta }}{15 k^2 \left(\eta _{\alpha \beta }+1\right) T_{\beta }^2 \left(\eta _{\alpha \beta }+\theta _{\alpha \beta }\right){}^2}\\
\mathcal{A}^{0231,1}_{\i\j}=\frac{32 \eta _{\alpha \beta }^3 m_{\beta }^2 n_{\alpha } n_{\beta }}{105 k^2 \left(\eta _{\alpha \beta }+1\right) T_{\beta }^2 \left(\eta _{\alpha \beta }+\theta _{\alpha \beta }\right){}^2}
\end{multline}

\begin{multline}
p=2,q=0:\    \mathcal{A}^{2011,1}_{\i\j} \Omega_{\i\j}^{11}+\mathcal{A}^{2021,1}_{\i\j} \Omega_{\i\j}^{12}+\mathcal{A}^{2031,1}_{\i\j} \Omega_{\i\j}^{13}+\mathcal{A}^{2022,1}_{\i\j} \Omega_{\i\j}^{22}+\mathcal{A}^{2032,1}_{\i\j} \Omega_{\i\j}^{23}+\mathcal{A}^{2033,1}_{\i\j} \Omega_{\i\j}^{33}\\
\mathcal{A}^{2011,1}_{\i\j}=\frac{140 k^2 \eta _{\alpha \beta } \theta _{\alpha \beta } n_{\alpha } n_{\beta } T_{\alpha }^2 \left(\eta _{\alpha \beta } \left(5 \theta _{\alpha \beta }-4\right)+\theta _{\alpha \beta }\right)}{3 \left(\eta
   _{\alpha \beta }+1\right){}^2 m_{\alpha }^2 \left(\eta _{\alpha \beta }+\theta _{\alpha \beta }\right){}^2}\\
\mathcal{A}^{2021,1}_{\i\j}=-\frac{112 k^2 \eta _{\alpha \beta } n_{\alpha } n_{\beta } T_{\alpha }^2 \left(\eta _{\alpha \beta }^3 \left(5 \theta _{\alpha \beta }^3-12 \theta _{\alpha \beta }^2+12 \theta _{\alpha \beta }-4\right)+3 \eta
   _{\alpha \beta }^2 \theta _{\alpha \beta }^3+3 \eta _{\alpha \beta } \theta _{\alpha \beta }^3+\theta _{\alpha \beta }^3\right)}{3 \left(\eta _{\alpha \beta }+1\right){}^4 \theta _{\alpha \beta } m_{\alpha }^2
   \left(\eta _{\alpha \beta }+\theta _{\alpha \beta }\right){}^2}\\
\mathcal{A}^{2031,1}_{\i\j}=\frac{16 k^2 \eta _{\alpha \beta } n_{\alpha } n_{\beta } T_{\alpha }^2 \left(\eta _{\alpha \beta }^2 \left(\theta _{\alpha \beta }^2-2 \theta _{\alpha \beta }+2\right)+2 \eta _{\alpha \beta } \theta _{\alpha
   \beta }+\theta _{\alpha \beta }^2\right) \left(\eta _{\alpha \beta }^2 \left(5 \theta _{\alpha \beta }^2-10 \theta _{\alpha \beta }+6\right)+2 \eta _{\alpha \beta } \theta _{\alpha \beta }+\theta _{\alpha
   \beta }^2\right)}{3 \left(\eta _{\alpha \beta }+1\right){}^5 \theta _{\alpha \beta }^2 m_{\alpha }^2 \left(\eta _{\alpha \beta }+\theta _{\alpha \beta }\right){}^2}\\
\mathcal{A}^{2022,1}_{\i\j}=-\frac{224 k^2 \eta _{\alpha \beta }^2 \left(\theta _{\alpha \beta }-1\right) n_{\alpha } n_{\beta } T_{\alpha }^2 \left(\eta _{\alpha \beta } \left(2 \theta _{\alpha \beta }-1\right)+\theta _{\alpha \beta
   }\right)}{3 \left(\eta _{\alpha \beta }+1\right){}^4 \theta _{\alpha \beta } m_{\alpha }^2 \left(\eta _{\alpha \beta }+\theta _{\alpha \beta }\right)}\\
\mathcal{A}^{2032,1}_{\i\j}=\frac{64 k^2 \eta _{\alpha \beta }^2 \left(\theta _{\alpha \beta }-1\right) n_{\alpha } n_{\beta } T_{\alpha }^2 \left(\eta _{\alpha \beta }^2 \left(2 \theta _{\alpha \beta }^2-4 \theta _{\alpha \beta
   }+3\right)+2 \eta _{\alpha \beta } \theta _{\alpha \beta }+\theta _{\alpha \beta }^2\right)}{3 \left(\eta _{\alpha \beta }+1\right){}^5 \theta _{\alpha \beta }^2 m_{\alpha }^2 \left(\eta _{\alpha \beta
   }+\theta _{\alpha \beta }\right)}\\
\mathcal{A}^{2033,1}_{\i\j}=\frac{64 k^2 \eta _{\alpha \beta }^3 \left(\theta _{\alpha \beta }-1\right){}^2 n_{\alpha } n_{\beta } T_{\alpha }^2}{3 \left(\eta _{\alpha \beta }+1\right){}^5 \theta _{\alpha \beta }^2 m_{\alpha }^2}
\end{multline}

\begin{multline}
p=1,q=2:\    \mathcal{A}^{1211,1}_{\i\j}\Omega_{\i\j}^{11}+\mathcal{A}^{1221,1}_{\i\j}\Omega_{\i\j}^{12}+\mathcal{A}^{1231,1}_{\i\j}\Omega_{\i\j}^{13}+\mathcal{A}^{1241,1}_{\i\j}\Omega_{\i\j}^{14}+\mathcal{A}^{1222,1}_{\i\j}\Omega_{\i\j}^{22}+\mathcal{A}^{1232,1}_{\i\j}\Omega_{\i\j}^{23}+\mathcal{A}^{1242,1}_{\i\j}\Omega_{\i\j}^{24}\\
\mathcal{A}^{1211,1}_{\i\j}=-\frac{4 \eta _{\alpha \beta }^2 \theta _{\alpha \beta } m_{\beta } n_{\alpha } n_{\beta } \left(\eta _{\alpha \beta } \left(27 \theta _{\alpha \beta }-10\right)+17 \theta _{\alpha \beta }\right)}{3 k \left(\eta
   _{\alpha \beta }+1\right){}^2 T_{\beta } \left(\eta _{\alpha \beta }+\theta _{\alpha \beta }\right){}^3}\\
\mathcal{A}^{1221,1}_{\i\j}=\frac{8 \eta _{\alpha \beta }^2 m_{\beta } n_{\alpha } n_{\beta } \left(\eta _{\alpha \beta }^2 \left(81 \theta _{\alpha \beta }^2-74 \theta _{\alpha \beta }+20\right)+2 \eta _{\alpha \beta } \theta _{\alpha
   \beta } \left(44 \theta _{\alpha \beta }-17\right)+27 \theta _{\alpha \beta }^2\right)}{15 k \left(\eta _{\alpha \beta }+1\right){}^3 T_{\beta } \left(\eta _{\alpha \beta }+\theta _{\alpha \beta }\right){}^3}\\
\mathcal{A}^{1231,1}_{\i\j}= -\frac{16 \eta _{\alpha \beta }^2 m_{\beta } n_{\alpha } n_{\beta } \left(\eta _{\alpha \beta }^2 \left(81 \theta _{\alpha \beta }^2-118 \theta _{\alpha \beta }+56\right)+2 \eta _{\alpha \beta } \theta _{\alpha
   \beta } \left(22 \theta _{\alpha \beta }-3\right)+19 \theta _{\alpha \beta }^2\right)}{105 k \left(\eta _{\alpha \beta }+1\right){}^3 T_{\beta } \left(\eta _{\alpha \beta }+\theta _{\alpha \beta }\right){}^3}\\
\mathcal{A}^{1241,1}_{\i\j}=\frac{32 \eta _{\alpha \beta }^2 m_{\beta } n_{\alpha } n_{\beta } \left(\eta _{\alpha \beta }^2 \left(3 \theta _{\alpha \beta }^2-6 \theta _{\alpha \beta }+4\right)+2 \eta _{\alpha \beta } \theta _{\alpha \beta
   }+\theta _{\alpha \beta }^2\right)}{105 k \left(\eta _{\alpha \beta }+1\right){}^3 T_{\beta } \left(\eta _{\alpha \beta }+\theta _{\alpha \beta }\right){}^3}\\
 \mathcal{A}^{1222,1}_{\i\j}=\frac{16 \eta _{\alpha \beta }^2 m_{\beta } n_{\alpha } n_{\beta } \left(\eta _{\alpha \beta } \left(9 \theta _{\alpha \beta }-5\right)+4 \theta _{\alpha \beta }\right)}{15 k \left(\eta _{\alpha \beta
   }+1\right){}^3 T_{\beta } \left(\eta _{\alpha \beta }+\theta _{\alpha \beta }\right){}^2}\\
  \mathcal{A}^{1232,1}_{\i\j}=-\frac{64 \eta _{\alpha \beta }^2 m_{\beta } n_{\alpha } n_{\beta } \left(\eta _{\alpha \beta } \left(9 \theta _{\alpha \beta }-7\right)+2 \theta _{\alpha \beta }\right)}{105 k \left(\eta _{\alpha \beta
   }+1\right){}^3 T_{\beta } \left(\eta _{\alpha \beta }+\theta _{\alpha \beta }\right){}^2}\\
   \mathcal{A}^{1242,1}_{\i\j}=\frac{64 \eta _{\alpha \beta }^3 \left(\theta _{\alpha \beta }-1\right) m_{\beta } n_{\alpha } n_{\beta }}{105 k \left(\eta _{\alpha \beta }+1\right){}^3 T_{\beta } \left(\eta _{\alpha \beta }+\theta _{\alpha
   \beta }\right){}^2}
\end{multline}

\begin{multline}
p=2,q=1:\    \mathcal{A}^{2111,1}_{\i\j}\Omega_{\i\j}^{11}+\mathcal{A}^{2121,1}_{\i\j}\Omega_{\i\j}^{12}+\mathcal{A}^{2131,1}_{\i\j}\Omega_{\i\j}^{13}+\mathcal{A}^{2141,1}_{\i\j}\Omega_{\i\j}^{14}+\mathcal{A}^{2122,1}_{\i\j}\Omega_{\i\j}^{22}+\mathcal{A}^{2132,1}_{\i\j}\Omega_{\i\j}^{23}+\mathcal{A}^{2142,1}_{\i\j}\Omega_{\i\j}^{24}\\+\mathcal{A}^{2133,1}_{\i\j}\Omega_{\i\j}^{33}+\mathcal{A}^{2143,1}_{\i\j}\Omega_{\i\j}^{34}\\
\mathcal{A}^{2111,1}_{\i\j}=-\frac{28 k \eta _{\alpha \beta } \theta _{\alpha \beta }^2 n_{\alpha } n_{\beta } T_{\alpha } \left(\eta _{\alpha \beta } \left(45 \theta _{\alpha \beta }-28\right)+17 \theta _{\alpha \beta }\right)}{3
   \left(\eta _{\alpha \beta }+1\right){}^2 m_{\alpha } \left(\eta _{\alpha \beta }+\theta _{\alpha \beta }\right){}^3}\\
\mathcal{A}^{2121,1}_{\i\j}= \frac{56 k \eta _{\alpha \beta } n_{\alpha } n_{\beta } T_{\alpha }}{15 \left(\eta _{\alpha \beta }+1\right){}^4 m_{\alpha } \left(\eta _{\alpha \beta }+\theta _{\alpha \beta }\right){}^3} \left(\eta _{\alpha \beta }^3 \left(135 \theta _{\alpha \beta }^3-236 \theta _{\alpha \beta }^2+168 \theta _{\alpha \beta }-40\right)+\eta
   _{\alpha \beta }^2 \theta _{\alpha \beta } \left(169 \theta _{\alpha \beta }^2-136 \theta _{\alpha \beta }+48\right)\right.\\
   \left.+\eta _{\alpha \beta } \theta _{\alpha \beta }^2 \left(101 \theta _{\alpha \beta
   }-20\right)+27 \theta _{\alpha \beta }^3\right)\\   
 \mathcal{A}^{2131,1}_{\i\j}=-\frac{16 k \eta _{\alpha \beta } n_{\alpha } n_{\beta } T_{\alpha }}{15 \left(\eta _{\alpha \beta }+1\right){}^5 \theta _{\alpha \beta } m_{\alpha } \left(\eta _{\alpha \beta }+\theta _{\alpha \beta }\right){}^3} \left(\eta _{\alpha \beta }^4 \left(135 \theta _{\alpha \beta }^4-388 \theta _{\alpha \beta }^3+492 \theta _{\alpha \beta }^2-280 \theta
   _{\alpha \beta }+60\right)\right.\\
   \left.+4 \eta _{\alpha \beta }^3 \theta _{\alpha \beta } \left(38 \theta _{\alpha \beta }^3-45 \theta _{\alpha \beta }^2+36 \theta _{\alpha \beta }-10\right)+6 \eta _{\alpha \beta }^2
   \theta _{\alpha \beta }^2 \left(23 \theta _{\alpha \beta }^2-6 \theta _{\alpha \beta }+2\right)+4 \eta _{\alpha \beta } \theta _{\alpha \beta }^3 \left(20 \theta _{\alpha \beta }-1\right)+19 \theta _{\alpha
   \beta }^4\right)\\
 \mathcal{A}^{2141,1}_{\i\j}=\frac{32 k \eta _{\alpha \beta } n_{\alpha } n_{\beta } T_{\alpha }}{{15 \left(\eta _{\alpha \beta }+1\right){}^5 \theta _{\alpha \beta }
   m_{\alpha } \left(\eta _{\alpha \beta }+\theta _{\alpha \beta }\right){}^3}} \left(\eta _{\alpha \beta }^4 \left(5 \theta _{\alpha \beta }^4-20 \theta _{\alpha \beta }^3+36 \theta _{\alpha \beta }^2-32 \theta _{\alpha
   \beta }+12\right)+4 \eta _{\alpha \beta }^3 \theta _{\alpha \beta } \left(3 \theta _{\alpha \beta }^2-6 \theta _{\alpha \beta }+4\right)\right.\\
   \left.+6 \eta _{\alpha \beta }^2 \theta _{\alpha \beta }^2 \left(\theta
   _{\alpha \beta }^2-2 \theta _{\alpha \beta }+2\right)+4 \eta _{\alpha \beta } \theta _{\alpha \beta }^3+\theta _{\alpha \beta }^4\right)\\
  \mathcal{A}^{2122,1}_{\i\j}= \frac{224 k \eta _{\alpha \beta } n_{\alpha } n_{\beta } T_{\alpha } \left(\eta _{\alpha \beta }^2 \left(18 \theta _{\alpha \beta }^2-21 \theta _{\alpha \beta }+5\right)+\eta _{\alpha \beta } \theta _{\alpha
   \beta } \left(15 \theta _{\alpha \beta }-11\right)+2 \theta _{\alpha \beta }^2\right)}{15 \left(\eta _{\alpha \beta }+1\right){}^4 m_{\alpha } \left(\eta _{\alpha \beta }+\theta _{\alpha \beta }\right){}^2}\\
   \mathcal{A}^{2132,1}_{\i\j}=-\frac{64 k \eta _{\alpha \beta } n_{\alpha } n_{\beta } T_{\alpha } \left(\eta _{\alpha \beta }^3 \left(36 \theta _{\alpha \beta }^3-75 \theta _{\alpha \beta }^2+56 \theta _{\alpha \beta }-15\right)+\eta
   _{\alpha \beta }^2 \theta _{\alpha \beta } \left(33 \theta _{\alpha \beta }^2-38 \theta _{\alpha \beta }+11\right)+2 \eta _{\alpha \beta } \theta _{\alpha \beta }^2 \left(7 \theta _{\alpha \beta }-4\right)+2
   \theta _{\alpha \beta }^3\right)}{15 \left(\eta _{\alpha \beta }+1\right){}^5 \theta _{\alpha \beta } m_{\alpha } \left(\eta _{\alpha \beta }+\theta _{\alpha \beta }\right){}^2}\\
   \mathcal{A}^{2142,1}_{\i\j}=\frac{128 k \eta _{\alpha \beta }^2 \left(\theta _{\alpha \beta }-1\right) n_{\alpha } n_{\beta } T_{\alpha } \left(\eta _{\alpha \beta }^2 \left(2 \theta _{\alpha \beta }^2-4 \theta _{\alpha \beta }+3\right)+2
   \eta _{\alpha \beta } \theta _{\alpha \beta }+\theta _{\alpha \beta }^2\right)}{15 \left(\eta _{\alpha \beta }+1\right){}^5 \theta _{\alpha \beta } m_{\alpha } \left(\eta _{\alpha \beta }+\theta _{\alpha \beta
   }\right){}^2}\\
  \mathcal{A}^{2133,1}_{\i\j} =-\frac{64 k \eta _{\alpha \beta }^2 \left(\theta _{\alpha \beta }-1\right) n_{\alpha } n_{\beta } T_{\alpha } \left(\eta _{\alpha \beta } \left(9 \theta _{\alpha \beta }-5\right)+4 \theta _{\alpha \beta
   }\right)}{15 \left(\eta _{\alpha \beta }+1\right){}^5 \theta _{\alpha \beta } m_{\alpha } \left(\eta _{\alpha \beta }+\theta _{\alpha \beta }\right)}\\
   \mathcal{A}^{2143,1}_{\i\j} =\frac{128 k \eta _{\alpha \beta }^3 \left(\theta _{\alpha \beta }-1\right){}^2 n_{\alpha } n_{\beta } T_{\alpha }}{15 \left(\eta _{\alpha \beta }+1\right){}^5 \theta _{\alpha \beta } m_{\alpha } \left(\eta
   _{\alpha \beta }+\theta _{\alpha \beta }\right)}
\end{multline}

\begin{multline}
p=2,q=2:\    \mathcal{A}^{2211,1}_{\i\j}\Omega_{\i\j}^{11}+\mathcal{A}^{2221,1}_{\i\j}\Omega_{\i\j}^{12}+\mathcal{A}^{2231,1}_{\i\j}\Omega_{\i\j}^{13}+\mathcal{A}^{2241,1}_{\i\j}\Omega_{\i\j}^{14}+\mathcal{A}^{2251,1}_{\i\j}\Omega_{\i\j}^{15}+\mathcal{A}^{2222,1}_{\i\j}\Omega_{\i\j}^{22}+\mathcal{A}^{2232,1}_{\i\j}\Omega_{\i\j}^{23}+\mathcal{A}^{2242,1}_{\i\j}\Omega_{\i\j}^{24}\\+\mathcal{A}^{2252,1}_{\i\j}\Omega_{\i\j}^{25}+\mathcal{A}^{2233,1}_{\i\j}\Omega_{\i\j}^{33}+\mathcal{A}^{2243,1}_{\i\j}\Omega_{\i\j}^{34}+\mathcal{A}^{2253,1}_{\i\j}\Omega_{\i\j}^{35}\\
\mathcal{A}^{2211,1}_{\i\j}=\frac{6 \eta _{\alpha \beta } \theta _{\alpha \beta }^3 n_{\alpha } n_{\beta } \left(\eta _{\alpha \beta } \left(55 \theta _{\alpha \beta }-28\right)+27 \theta _{\alpha \beta }\right)}{\left(\eta _{\alpha \beta
   }+1\right){}^2 \left(\eta _{\alpha \beta }+\theta _{\alpha \beta }\right){}^4}\\
\mathcal{A}^{2221,1}_{\i\j}=-\frac{16 \eta _{\alpha \beta } \theta _{\alpha \beta } n_{\alpha } n_{\beta }}{{15 \left(\eta _{\alpha \beta }+1\right){}^4 \left(\eta _{\alpha \beta }+\theta _{\alpha \beta }\right){}^4}} \left(\eta _{\alpha \beta }^3 \left(495 \theta _{\alpha \beta }^3-684 \theta _{\alpha \beta }^2+378 \theta _{\alpha \beta
   }-70\right)+3 \eta _{\alpha \beta }^2 \theta _{\alpha \beta } \left(267 \theta _{\alpha \beta }^2-204 \theta _{\alpha \beta }+56\right)\right.\\
   \left.+3 \eta _{\alpha \beta } \theta _{\alpha \beta }^2 \left(165 \theta
   _{\alpha \beta }-46\right)+119 \theta _{\alpha \beta }^3\right)\\
\mathcal{A}^{2231,1}_{\i\j}=\frac{16 \eta _{\alpha \beta } n_{\alpha } n_{\beta }}{105 \left(\eta _{\alpha \beta }+1\right){}^5 \left(\eta _{\alpha \beta }+\theta _{\alpha \beta }\right){}^4} \left(\eta _{\alpha \beta }^4 \left(1485 \theta _{\alpha \beta }^4-3348 \theta _{\alpha \beta }^3+3294 \theta _{\alpha \beta }^2-1400 \theta _{\alpha \beta
   }+210\right)\right.\\
   \left.+4 \eta _{\alpha \beta }^3 \theta _{\alpha \beta } \left(648 \theta _{\alpha \beta }^3-864 \theta _{\alpha \beta }^2+597 \theta _{\alpha \beta }-140\right)+6 \eta _{\alpha \beta }^2 \theta _{\alpha
   \beta }^2 \left(360 \theta _{\alpha \beta }^2-178 \theta _{\alpha \beta }+59\right)+4 \eta _{\alpha \beta } \theta _{\alpha \beta }^3 \left(271 \theta _{\alpha \beta }-30\right)+241 \theta _{\alpha \beta
   }^4\right)\\
\mathcal{A}^{2241,1}_{\i\j}=-\frac{64 \eta _{\alpha \beta } n_{\alpha } n_{\beta }}{105
   \left(\eta _{\alpha \beta }+1\right){}^5 \left(\eta _{\alpha \beta }+\theta _{\alpha \beta }\right){}^4} \left(\eta _{\alpha \beta }^4 \left(55 \theta _{\alpha \beta }^4-172 \theta _{\alpha \beta }^3+240 \theta _{\alpha \beta }^2-158 \theta _{\alpha \beta
   }+42\right)\right.\\
   \left.+2 \eta _{\alpha \beta }^3 \theta _{\alpha \beta } \left(24 \theta _{\alpha \beta }^3-18 \theta _{\alpha \beta }^2+3 \theta _{\alpha \beta }+5\right)+6 \eta _{\alpha \beta }^2 \theta _{\alpha \beta
   }^2 \left(9 \theta _{\alpha \beta }^2-5 \theta _{\alpha \beta }+3\right)+2 \eta _{\alpha \beta } \theta _{\alpha \beta }^3 \left(13 \theta _{\alpha \beta }+1\right)+7 \theta _{\alpha \beta }^4\right)\\
\mathcal{A}^{2251,1}_{\i\j}=\frac{32 \eta _{\alpha \beta } n_{\alpha } n_{\beta }}{105 \left(\eta _{\alpha \beta }+1\right){}^5 \left(\eta _{\alpha \beta }+\theta
   _{\alpha \beta }\right){}^4} \left(\eta _{\alpha \beta }^4 \left(5 \theta _{\alpha \beta }^4-20 \theta _{\alpha \beta }^3+36 \theta _{\alpha \beta }^2-32 \theta _{\alpha \beta
   }+12\right)+4 \eta _{\alpha \beta }^3 \theta _{\alpha \beta } \left(3 \theta _{\alpha \beta }^2-6 \theta _{\alpha \beta }+4\right)\right.\\
   \left.+6 \eta _{\alpha \beta }^2 \theta _{\alpha \beta }^2 \left(\theta _{\alpha
   \beta }^2-2 \theta _{\alpha \beta }+2\right)+4 \eta _{\alpha \beta } \theta _{\alpha \beta }^3+\theta _{\alpha \beta }^4\right)\\
\mathcal{A}^{2222,1}_{\i\j}=-\frac{16 \eta _{\alpha \beta } \theta _{\alpha \beta } n_{\alpha } n_{\beta } \left(\eta _{\alpha \beta }^2 \left(198 \theta _{\alpha \beta }^2-189 \theta _{\alpha \beta }+35\right)+\eta _{\alpha \beta } \theta
   _{\alpha \beta } \left(207 \theta _{\alpha \beta }-119\right)+44 \theta _{\alpha \beta }^2\right)}{15 \left(\eta _{\alpha \beta }+1\right){}^4 \left(\eta _{\alpha \beta }+\theta _{\alpha \beta }\right){}^3}\\
\mathcal{A}^{2232,1}_{\i\j}=\frac{32 \eta _{\alpha \beta } n_{\alpha } n_{\beta }}{105 \left(\eta _{\alpha \beta }+1\right){}^5 \left(\eta _{\alpha \beta }+\theta _{\alpha \beta }\right){}^3} \left(\eta _{\alpha \beta }^3 \left(594 \theta _{\alpha \beta }^3-972 \theta _{\alpha \beta }^2+539 \theta _{\alpha \beta }-105\right)+2 \eta _{\alpha \beta
   }^2 \theta _{\alpha \beta } \left(405 \theta _{\alpha \beta }^2-433 \theta _{\alpha \beta }+112\right)\right.\\
   \left.+\eta _{\alpha \beta } \theta _{\alpha \beta }^2 \left(377 \theta _{\alpha \beta }-209\right)+56 \theta
   _{\alpha \beta }^3\right)\\
\mathcal{A}^{2242,1}_{\i\j}=-\frac{64 \eta _{\alpha \beta } n_{\alpha } n_{\beta } \left(\eta _{\alpha \beta }^3 \left(66 \theta _{\alpha \beta }^3-153 \theta _{\alpha \beta }^2+133 \theta _{\alpha \beta }-42\right)+\eta _{\alpha \beta }^2
   \theta _{\alpha \beta } \left(45 \theta _{\alpha \beta }^2-40 \theta _{\alpha \beta }+7\right)+\eta _{\alpha \beta } \theta _{\alpha \beta }^2 \left(25 \theta _{\alpha \beta }-13\right)+4 \theta _{\alpha \beta
   }^3\right)}{105 \left(\eta _{\alpha \beta }+1\right){}^5 \left(\eta _{\alpha \beta }+\theta _{\alpha \beta }\right){}^3}\\
\mathcal{A}^{2252,1}_{\i\j}=\frac{128 \eta _{\alpha \beta }^2 \left(\theta _{\alpha \beta }-1\right) n_{\alpha } n_{\beta } \left(\eta _{\alpha \beta }^2 \left(2 \theta _{\alpha \beta }^2-4 \theta _{\alpha \beta }+3\right)+2 \eta _{\alpha
   \beta } \theta _{\alpha \beta }+\theta _{\alpha \beta }^2\right)}{105 \left(\eta _{\alpha \beta }+1\right){}^5 \left(\eta _{\alpha \beta }+\theta _{\alpha \beta }\right){}^3}\\   
\mathcal{A}^{2233,1}_{\i\j}=\frac{32 \eta _{\alpha \beta } n_{\alpha } n_{\beta } \left(\eta _{\alpha \beta }^2 \left(99 \theta _{\alpha \beta }^2-126 \theta _{\alpha \beta }+35\right)+8 \eta _{\alpha \beta } \theta _{\alpha \beta } \left(9
   \theta _{\alpha \beta }-7\right)+8 \theta _{\alpha \beta }^2\right)}{105 \left(\eta _{\alpha \beta }+1\right){}^5 \left(\eta _{\alpha \beta }+\theta _{\alpha \beta }\right){}^2}\\
\mathcal{A}^{2243,1}_{\i\j}=-\frac{128 \eta _{\alpha \beta }^2 \left(\theta _{\alpha \beta }-1\right) n_{\alpha } n_{\beta } \left(\eta _{\alpha \beta } \left(11 \theta _{\alpha \beta }-7\right)+4 \theta _{\alpha \beta }\right)}{105
   \left(\eta _{\alpha \beta }+1\right){}^5 \left(\eta _{\alpha \beta }+\theta _{\alpha \beta }\right){}^2}\\
\mathcal{A}^{2253,1}_{\i\j}=\frac{128 \eta _{\alpha \beta }^3 \left(\theta _{\alpha \beta }-1\right){}^2 n_{\alpha } n_{\beta }}{105 \left(\eta _{\alpha \beta }+1\right){}^5 \left(\eta _{\alpha \beta }+\theta _{\alpha \beta }\right){}^2}
\end{multline}

\subsection{Values of $A^{1pq}_{\i\j}$ for $p,q\leq2$}

\begin{multline}
p=0,q=0:\    \mathcal{A}^{0011,1}_{\i\i}\Omega_{\i\j}^{11}\\
\mathcal{A}^{0011,1}_{\i\i}=-\frac{16 n_{\alpha } n_{\beta }}{3 \eta _{\alpha \beta }+3}
\end{multline}

\begin{multline}
p=0,q=1:\    \mathcal{A}^{0111,1}_{\i\i}\Omega_{\i\j}^{11}+\mathcal{A}^{0121,1}_{\i\i}\Omega_{\i\j}^{12}\\
\mathcal{A}^{0111,1}_{\i\i}=\frac{16 \theta _{\alpha \beta } m_{\alpha } n_{\alpha } n_{\beta }}{3 k \left(\eta _{\alpha \beta }+1\right) T_{\alpha } \left(\eta _{\alpha \beta }+\theta _{\alpha \beta }\right)}\\
\mathcal{A}^{0121,1}_{\i\i}=-\frac{32 \theta _{\alpha \beta } m_{\alpha } n_{\alpha } n_{\beta }}{15 k \left(\eta _{\alpha \beta }+1\right) T_{\alpha } \left(\eta _{\alpha \beta }+\theta _{\alpha \beta }\right)}
\end{multline}

\begin{multline}
p=1,q=0:\    \mathcal{A}^{1011,1}_{\i\i}\Omega_{\i\j}^{11}+\mathcal{A}^{1021,1}_{\i\i}\Omega_{\i\j}^{12}+\mathcal{A}^{1022,1}_{\i\i}\Omega_{\i\j}^{22}\\
\mathcal{A}^{1011,1}_{\i\i}=\frac{40 k n_{\alpha } n_{\beta } T_{\alpha } \left(-2 \eta _{\alpha \beta }^2 \left(\theta _{\alpha \beta }-1\right)+\eta _{\alpha \beta } \theta _{\alpha \beta }^2+\theta _{\alpha \beta }^2\right)}{3 \left(\eta
   _{\alpha \beta }+1\right){}^2 \theta _{\alpha \beta } m_{\alpha } \left(\eta _{\alpha \beta }+\theta _{\alpha \beta }\right)}\\
\mathcal{A}^{1021,1}_{\i\i}=-\frac{16 k n_{\alpha } n_{\beta } T_{\alpha } \left(\eta _{\alpha \beta }^2 \left(3 \theta _{\alpha \beta }^2-6 \theta _{\alpha \beta }+4\right)+2 \eta _{\alpha \beta } \theta _{\alpha \beta }+\theta _{\alpha
   \beta }^2\right)}{3 \left(\eta _{\alpha \beta }+1\right){}^3 \theta _{\alpha \beta } m_{\alpha } \left(\eta _{\alpha \beta }+\theta _{\alpha \beta }\right)}\\
\mathcal{A}^{1022,1}_{\i\i}=-\frac{32 k \eta _{\alpha \beta } \left(\theta _{\alpha \beta }-1\right) n_{\alpha } n_{\beta } T_{\alpha }}{3 \left(\eta _{\alpha \beta }+1\right){}^3 \theta _{\alpha \beta } m_{\alpha }} 
\end{multline}

\begin{multline}
p=1,q=1:\    \mathcal{A}^{1111,1}_{\i\i}\Omega_{\i\j}^{11}+\mathcal{A}^{1121,1}_{\i\i}\Omega_{\i\j}^{12}+\mathcal{A}^{1131,1}_{\i\i}\Omega_{\i\j}^{13}+\mathcal{A}^{1122,1}_{\i\i}\Omega_{\i\j}^{22}+\mathcal{A}^{1132,1}_{\i\i}\Omega_{\i\j}^{23}\\
\mathcal{A}^{1111,1}_{\i\i}=-\frac{8 n_{\alpha } n_{\beta } \left(-2 \eta _{\alpha \beta }^2 \left(5 \theta _{\alpha \beta }-8\right)+5 \eta _{\alpha \beta } \theta _{\alpha \beta }^2+6 \eta _{\alpha \beta }^3+5 \theta _{\alpha \beta
   }^2\right)}{3 \left(\eta _{\alpha \beta }+1\right){}^2 \left(\eta _{\alpha \beta }+\theta _{\alpha \beta }\right){}^2}\\
\mathcal{A}^{1121,1}_{\i\i}=-\frac{32 n_{\alpha } n_{\beta } \left(11 \eta _{\alpha \beta }^3 \left(\theta _{\alpha \beta }-1\right)+\eta _{\alpha \beta }^2 \left(-10 \theta _{\alpha \beta }^2+26 \theta _{\alpha \beta }-21\right)-5 \eta
   _{\alpha \beta } \theta _{\alpha \beta } \left(\theta _{\alpha \beta }+1\right)-5 \theta _{\alpha \beta }^2\right)}{15 \left(\eta _{\alpha \beta }+1\right){}^3 \left(\eta _{\alpha \beta }+\theta _{\alpha \beta
   }\right){}^2}\\
\mathcal{A}^{1131,1}_{\i\i}=-\frac{32 n_{\alpha } n_{\beta } \left(\eta _{\alpha \beta }^2 \left(3 \theta _{\alpha \beta }^2-6 \theta _{\alpha \beta }+4\right)+2 \eta _{\alpha \beta } \theta _{\alpha \beta }+\theta _{\alpha \beta
   }^2\right)}{15 \left(\eta _{\alpha \beta }+1\right){}^3 \left(\eta _{\alpha \beta }+\theta _{\alpha \beta }\right){}^2}\\
\mathcal{A}^{1122,1}_{\i\i}=-\frac{32 \eta _{\alpha \beta } n_{\alpha } n_{\beta } \left(2 \eta _{\alpha \beta }-5 \theta _{\alpha \beta }+7\right)}{15 \left(\eta _{\alpha \beta }+1\right){}^3 \left(\eta _{\alpha \beta }+\theta _{\alpha
   \beta }\right)}\\
\mathcal{A}^{1132,1}_{\i\i}=-\frac{64 \eta _{\alpha \beta } \left(\theta _{\alpha \beta }-1\right) n_{\alpha } n_{\beta }}{15 \left(\eta _{\alpha \beta }+1\right){}^3 \left(\eta _{\alpha \beta }+\theta _{\alpha \beta }\right)} 
\end{multline}

\begin{multline}
p=0,q=2:\    \mathcal{A}^{0211,1}_{\i\i}\Omega_{\i\j}^{11}+\mathcal{A}^{0221,1}_{\i\i}\Omega_{\i\j}^{12}+\mathcal{A}^{0231,1}_{\i\i}\Omega_{\i\j}^{13}\\
\mathcal{A}^{0211,1}_{\i\i}=-\frac{8 \theta _{\alpha \beta }^2 m_{\alpha }^2 n_{\alpha } n_{\beta }}{3 k^2 \left(\eta _{\alpha \beta }+1\right) T_{\alpha }^2 \left(\eta _{\alpha \beta }+\theta _{\alpha \beta }\right){}^2}\\
\mathcal{A}^{0221,1}_{\i\i}=\frac{32 \theta _{\alpha \beta }^2 m_{\alpha }^2 n_{\alpha } n_{\beta }}{15 k^2 \left(\eta _{\alpha \beta }+1\right) T_{\alpha }^2 \left(\eta _{\alpha \beta }+\theta _{\alpha \beta }\right){}^2}\\
\mathcal{A}^{0231,1}_{\i\i}=-\frac{32 \theta _{\alpha \beta }^2 m_{\alpha }^2 n_{\alpha } n_{\beta }}{105 k^2 \left(\eta _{\alpha \beta }+1\right) T_{\alpha }^2 \left(\eta _{\alpha \beta }+\theta _{\alpha \beta }\right){}^2}  
\end{multline}

\begin{multline}
p=2,q=0:\    \mathcal{A}^{2011,1}_{\i\i}\Omega_{\i\j}^{11}+\mathcal{A}^{2021,1}_{\i\i}\Omega_{\i\j}^{12}+\mathcal{A}^{2031,1}_{\i\i}\Omega_{\i\j}^{13}+\mathcal{A}^{2022,1}_{\i\i}\Omega_{\i\j}^{22}+\mathcal{A}^{2032,1}_{\i\i}\Omega_{\i\j}^{23}++\mathcal{A}^{2033,1}_{\i\i}\Omega_{\i\j}^{33}\\
\mathcal{A}^{2011,1}_{\i\i}=\frac{140 k^2 n_{\alpha } n_{\beta } T_{\alpha }^2 \left(4 \eta _{\alpha \beta }^2 \left(\theta _{\alpha \beta }-1\right)-\eta _{\alpha \beta } \theta _{\alpha \beta }^2-\theta _{\alpha \beta }^2\right)}{3
   \left(\eta _{\alpha \beta }+1\right){}^2 m_{\alpha }^2 \left(\eta _{\alpha \beta }+\theta _{\alpha \beta }\right){}^2}\\
\mathcal{A}^{2021,1}_{\i\i}=\frac{112 k^2 n_{\alpha } n_{\beta } T_{\alpha }^2}{3 \left(\eta _{\alpha \beta }+1\right){}^4 \theta _{\alpha \beta }^2
   m_{\alpha }^2 \left(\eta _{\alpha \beta }+\theta _{\alpha \beta }\right){}^2} \left(-2 \eta _{\alpha \beta }^4 \left(\theta _{\alpha \beta }^3-3 \theta _{\alpha \beta }^2+4 \theta _{\alpha \beta }-2\right)+\eta _{\alpha \beta }^3 \theta
   _{\alpha \beta } \left(3 \theta _{\alpha \beta }^3-6 \theta _{\alpha \beta }^2+4\right)\right.\\
   \left.+3 \eta _{\alpha \beta }^2 \theta _{\alpha \beta }^2 \left(\theta _{\alpha \beta }^2-2 \theta _{\alpha \beta
   }+2\right)+\eta _{\alpha \beta } \theta _{\alpha \beta }^3 \left(\theta _{\alpha \beta }+2\right)+\theta _{\alpha \beta }^4\right)\\   
\mathcal{A}^{2031,1}_{\i\i}=-\frac{16 k^2 n_{\alpha } n_{\beta } T_{\alpha }^2}{3 \left(\eta _{\alpha \beta }+1\right){}^5 \theta _{\alpha \beta }^2 m_{\alpha }^2 \left(\eta
   _{\alpha \beta }+\theta _{\alpha \beta }\right){}^2} \left(\eta _{\alpha \beta }^4 \left(5 \theta _{\alpha \beta }^4-20 \theta _{\alpha \beta }^3+36 \theta _{\alpha \beta }^2-32 \theta _{\alpha \beta }+12\right)\right.\\
   \left.+4
   \eta _{\alpha \beta }^3 \theta _{\alpha \beta } \left(3 \theta _{\alpha \beta }^2-6 \theta _{\alpha \beta }+4\right)+6 \eta _{\alpha \beta }^2 \theta _{\alpha \beta }^2 \left(\theta _{\alpha \beta }^2-2 \theta
   _{\alpha \beta }+2\right)+4 \eta _{\alpha \beta } \theta _{\alpha \beta }^3+\theta _{\alpha \beta }^4\right)\\
\mathcal{A}^{2022,1}_{\i\i}=\frac{224 k^2 \eta _{\alpha \beta } \left(\theta _{\alpha \beta }-1\right) n_{\alpha } n_{\beta } T_{\alpha }^2 \left(\eta _{\alpha \beta }^2 \left(-\left(\theta _{\alpha \beta }-1\right)\right)+\eta _{\alpha
   \beta } \theta _{\alpha \beta }^2+\theta _{\alpha \beta }^2\right)}{3 \left(\eta _{\alpha \beta }+1\right){}^4 \theta _{\alpha \beta }^2 m_{\alpha }^2 \left(\eta _{\alpha \beta }+\theta _{\alpha \beta
   }\right)}\\
\mathcal{A}^{2032,1}_{\i\i}= -\frac{64 k^2 \eta _{\alpha \beta } \left(\theta _{\alpha \beta }-1\right) n_{\alpha } n_{\beta } T_{\alpha }^2 \left(\eta _{\alpha \beta }^2 \left(2 \theta _{\alpha \beta }^2-4 \theta _{\alpha \beta }+3\right)+2
   \eta _{\alpha \beta } \theta _{\alpha \beta }+\theta _{\alpha \beta }^2\right)}{3 \left(\eta _{\alpha \beta }+1\right){}^5 \theta _{\alpha \beta }^2 m_{\alpha }^2 \left(\eta _{\alpha \beta }+\theta _{\alpha
   \beta }\right)}\\
\mathcal{A}^{2033,1}_{\i\i}=-\frac{64 k^2 \eta _{\alpha \beta }^2 \left(\theta _{\alpha \beta }-1\right){}^2 n_{\alpha } n_{\beta } T_{\alpha }^2}{3 \left(\eta _{\alpha \beta }+1\right){}^5 \theta _{\alpha \beta }^2 m_{\alpha }^2}
\end{multline}

\begin{multline}
p=1,q=2:\    \mathcal{A}^{1211,1}_{\i\i}\Omega_{\i\j}^{11}+\mathcal{A}^{1221,1}_{\i\i}\Omega_{\i\j}^{12}+\mathcal{A}^{1231,1}_{\i\i}\Omega_{\i\j}^{13}+\mathcal{A}^{1241,1}_{\i\i}\Omega_{\i\j}^{14}+\mathcal{A}^{1222,1}_{\i\i}\Omega_{\i\j}^{22}+\mathcal{A}^{1232,1}_{\i\i}\Omega_{\i\j}^{23}+\mathcal{A}^{1242,1}_{\i\i}\Omega_{\i\j}^{24}\\
\mathcal{A}^{1211,1}_{\i\i}=\frac{4 \theta _{\alpha \beta } m_{\alpha } n_{\alpha } n_{\beta } \left(\eta _{\alpha \beta }^2 \left(22-10 \theta _{\alpha \beta }\right)+5 \eta _{\alpha \beta } \theta _{\alpha \beta }^2+12 \eta _{\alpha \beta
   }^3+5 \theta _{\alpha \beta }^2\right)}{3 k \left(\eta _{\alpha \beta }+1\right){}^2 T_{\alpha } \left(\eta _{\alpha \beta }+\theta _{\alpha \beta }\right){}^3}\\
\mathcal{A}^{1221,1}_{\i\i}=-\frac{8 \theta _{\alpha \beta } m_{\alpha } n_{\alpha } n_{\beta } \left(\eta _{\alpha \beta }^3 \left(68-44 \theta _{\alpha \beta }\right)+\eta _{\alpha \beta }^2 \left(25 \theta _{\alpha \beta }^2-74 \theta
   _{\alpha \beta }+76\right)+10 \eta _{\alpha \beta } \theta _{\alpha \beta } \left(2 \theta _{\alpha \beta }+1\right)+12 \eta _{\alpha \beta }^4+15 \theta _{\alpha \beta }^2\right)}{15 k \left(\eta _{\alpha
   \beta }+1\right){}^3 T_{\alpha } \left(\eta _{\alpha \beta }+\theta _{\alpha \beta }\right){}^3}\\
\mathcal{A}^{1231,1}_{\i\i}=\frac{16 \theta _{\alpha \beta } m_{\alpha } n_{\alpha } n_{\beta } \left(-34 \eta _{\alpha \beta }^3 \left(\theta _{\alpha \beta }-1\right)+\eta _{\alpha \beta }^2 \left(47 \theta _{\alpha \beta }^2-118 \theta
   _{\alpha \beta }+90\right)+2 \eta _{\alpha \beta } \theta _{\alpha \beta } \left(5 \theta _{\alpha \beta }+14\right)+19 \theta _{\alpha \beta }^2\right)}{105 k \left(\eta _{\alpha \beta }+1\right){}^3
   T_{\alpha } \left(\eta _{\alpha \beta }+\theta _{\alpha \beta }\right){}^3}\\
\mathcal{A}^{1241,1}_{\i\i}=-\frac{32 \theta _{\alpha \beta } m_{\alpha } n_{\alpha } n_{\beta } \left(\eta _{\alpha \beta }^2 \left(3 \theta _{\alpha \beta }^2-6 \theta _{\alpha \beta }+4\right)+2 \eta _{\alpha \beta } \theta _{\alpha
   \beta }+\theta _{\alpha \beta }^2\right)}{105 k \left(\eta _{\alpha \beta }+1\right){}^3 T_{\alpha } \left(\eta _{\alpha \beta }+\theta _{\alpha \beta }\right){}^3}\\
\mathcal{A}^{1222,1}_{\i\i}=\frac{16 \eta _{\alpha \beta } \theta _{\alpha \beta } m_{\alpha } n_{\alpha } n_{\beta } \left(4 \eta _{\alpha \beta }-5 \theta _{\alpha \beta }+9\right)}{15 k \left(\eta _{\alpha \beta }+1\right){}^3 T_{\alpha
   } \left(\eta _{\alpha \beta }+\theta _{\alpha \beta }\right){}^2}\\
\mathcal{A}^{1232,1}_{\i\i}=-\frac{64 \eta _{\alpha \beta } \theta _{\alpha \beta } m_{\alpha } n_{\alpha } n_{\beta } \left(2 \eta _{\alpha \beta }-7 \theta _{\alpha \beta }+9\right)}{105 k \left(\eta _{\alpha \beta }+1\right){}^3
   T_{\alpha } \left(\eta _{\alpha \beta }+\theta _{\alpha \beta }\right){}^2}\\
\mathcal{A}^{1242,1}_{\i\i}=-\frac{64 \eta _{\alpha \beta } \left(\theta _{\alpha \beta }-1\right) \theta _{\alpha \beta } m_{\alpha } n_{\alpha } n_{\beta }}{105 k \left(\eta _{\alpha \beta }+1\right){}^3 T_{\alpha } \left(\eta _{\alpha
   \beta }+\theta _{\alpha \beta }\right){}^2}
\end{multline}

\begin{multline}
p=2,q=1:\    \mathcal{A}^{2111,1}_{\i\i}\Omega_{\i\j}^{11}+\mathcal{A}^{2121,1}_{\i\i}\Omega_{\i\j}^{12}+\mathcal{A}^{2131,1}_{\i\i}\Omega_{\i\j}^{13}+\mathcal{A}^{2141,1}_{\i\i}\Omega_{\i\j}^{14}+\mathcal{A}^{2122,1}_{\i\i}\Omega_{\i\j}^{22}+\mathcal{A}^{2132,1}_{\i\i}\Omega_{\i\j}^{23}\\
+\mathcal{A}^{2142,1}_{\i\i}\Omega_{\i\j}^{24}+\mathcal{A}^{2133,1}_{\i\i}\Omega_{\i\j}^{33}+\mathcal{A}^{2143,1}_{\i\i}\Omega_{\i\j}^{34}\\
\mathcal{A}^{2111,1}_{\i\i}=-\frac{28 k n_{\alpha } n_{\beta } T_{\alpha } \left(8 \eta _{\alpha \beta }^4 \left(\theta _{\alpha \beta }-1\right)-12 \eta _{\alpha \beta }^3 \theta _{\alpha \beta }^2+4 \eta _{\alpha \beta }^2 \theta _{\alpha
   \beta }^2 \left(5 \theta _{\alpha \beta }-8\right)-5 \eta _{\alpha \beta } \theta _{\alpha \beta }^4-5 \theta _{\alpha \beta }^4\right)}{3 \left(\eta _{\alpha \beta }+1\right){}^2 \theta _{\alpha \beta }
   m_{\alpha } \left(\eta _{\alpha \beta }+\theta _{\alpha \beta }\right){}^3}\\
\mathcal{A}^{2121,1}_{\i\i}=-\frac{56 k n_{\alpha } n_{\beta } T_{\alpha }}{15 \left(\eta _{\alpha \beta }+1\right){}^4 \theta _{\alpha \beta } m_{\alpha } \left(\eta _{\alpha \beta }+\theta _{\alpha \beta }\right){}^3} \left(12 \eta _{\alpha \beta }^5 \left(3 \theta _{\alpha \beta }^2-6 \theta _{\alpha \beta }+4\right)-4 \eta _{\alpha \beta }^4 \left(16 \theta _{\alpha \beta }^3-35
   \theta _{\alpha \beta }^2+32 \theta _{\alpha \beta }-22\right)\right.\\
   \left.+\eta _{\alpha \beta }^3 \theta _{\alpha \beta } \left(35 \theta _{\alpha \beta }^3-148 \theta _{\alpha \beta }^2+100 \theta _{\alpha \beta
   }+64\right)+\eta _{\alpha \beta }^2 \theta _{\alpha \beta }^2 \left(45 \theta _{\alpha \beta }^2-104 \theta _{\alpha \beta }+116\right)+5 \eta _{\alpha \beta } \theta _{\alpha \beta }^3 \left(5 \theta _{\alpha
   \beta }+4\right)+15 \theta _{\alpha \beta }^4\right)\\
\mathcal{A}^{2131,1}_{\i\i}=-\frac{16 k n_{\alpha } n_{\beta } T_{\alpha }}{15 \left(\eta _{\alpha \beta }+1\right){}^5 \theta _{\alpha \beta } m_{\alpha }
   \left(\eta _{\alpha \beta }+\theta _{\alpha \beta }\right){}^3} \left(4 \eta _{\alpha \beta }^5 \left(17 \theta _{\alpha \beta }^3-51 \theta _{\alpha \beta }^2+64 \theta _{\alpha \beta }-30\right)\right.\\
   \left.+\eta _{\alpha \beta }^4
   \left(-67 \theta _{\alpha \beta }^4+252 \theta _{\alpha \beta }^3-336 \theta _{\alpha \beta }^2+312 \theta _{\alpha \beta }-180\right)-4 \eta _{\alpha \beta }^3 \theta _{\alpha \beta } \left(21 \theta _{\alpha
   \beta }^3-33 \theta _{\alpha \beta }^2-15 \theta _{\alpha \beta }+46\right)\right.\\
   \left.-2 \eta _{\alpha \beta }^2 \theta _{\alpha \beta }^2 \left(43 \theta _{\alpha \beta }^2-70 \theta _{\alpha \beta }+84\right)-4 \eta
   _{\alpha \beta } \theta _{\alpha \beta }^3 \left(7 \theta _{\alpha \beta }+12\right)-19 \theta _{\alpha \beta }^4\right)\\
\mathcal{A}^{2141,1}_{\i\i}=-\frac{32 k n_{\alpha } n_{\beta } T_{\alpha }}{15 \left(\eta _{\alpha \beta }+1\right){}^5 \theta _{\alpha \beta } m_{\alpha } \left(\eta _{\alpha
   \beta }+\theta _{\alpha \beta }\right){}^3} \left(\eta _{\alpha \beta }^4 \left(5 \theta _{\alpha \beta }^4-20 \theta _{\alpha \beta }^3+36 \theta _{\alpha \beta }^2-32 \theta _{\alpha \beta }+12\right)\right.\\
   \left.+4 \eta
   _{\alpha \beta }^3 \theta _{\alpha \beta } \left(3 \theta _{\alpha \beta }^2-6 \theta _{\alpha \beta }+4\right)+6 \eta _{\alpha \beta }^2 \theta _{\alpha \beta }^2 \left(\theta _{\alpha \beta }^2-2 \theta
   _{\alpha \beta }+2\right)+4 \eta _{\alpha \beta } \theta _{\alpha \beta }^3+\theta _{\alpha \beta }^4\right)\\
\mathcal{A}^{2122,1}_{\i\i}=-\frac{224 k \eta _{\alpha \beta } n_{\alpha } n_{\beta } T_{\alpha } \left(6 \eta _{\alpha \beta }^3 \left(\theta _{\alpha \beta }-1\right)+\eta _{\alpha \beta }^2 \left(-7 \theta _{\alpha \beta }^2+16 \theta
   _{\alpha \beta }-11\right)+\eta _{\alpha \beta } \theta _{\alpha \beta }^2 \left(5 \theta _{\alpha \beta }-9\right)+\theta _{\alpha \beta }^2 \left(5 \theta _{\alpha \beta }-7\right)\right)}{15 \left(\eta
   _{\alpha \beta }+1\right){}^4 \theta _{\alpha \beta } m_{\alpha } \left(\eta _{\alpha \beta }+\theta _{\alpha \beta }\right){}^2}\\   
\mathcal{A}^{2132,1}_{\i\i}=-\frac{64 k \eta _{\alpha \beta } n_{\alpha } n_{\beta } T_{\alpha }}{15 \left(\eta _{\alpha \beta }+1\right){}^5 \theta _{\alpha \beta } m_{\alpha } \left(\eta _{\alpha \beta }+\theta _{\alpha \beta
   }\right){}^2} \left(\eta _{\alpha \beta }^3 \left(19 \theta _{\alpha \beta }^2-38 \theta _{\alpha \beta }+21\right)+\eta _{\alpha \beta }^2 \left(-17 \theta
   _{\alpha \beta }^3+56 \theta _{\alpha \beta }^2-69 \theta _{\alpha \beta }+36\right)\right.\\
   \left.+2 \eta _{\alpha \beta } \theta _{\alpha \beta } \left(-7 \theta _{\alpha \beta }^2+3 \theta _{\alpha \beta }+7\right)+2
   \left(7-6 \theta _{\alpha \beta }\right) \theta _{\alpha \beta }^2\right)\\
\mathcal{A}^{2142,1}_{\i\i}=-\frac{128 k \eta _{\alpha \beta } \left(\theta _{\alpha \beta }-1\right) n_{\alpha } n_{\beta } T_{\alpha } \left(\eta _{\alpha \beta }^2 \left(2 \theta _{\alpha \beta }^2-4 \theta _{\alpha \beta }+3\right)+2
   \eta _{\alpha \beta } \theta _{\alpha \beta }+\theta _{\alpha \beta }^2\right)}{15 \left(\eta _{\alpha \beta }+1\right){}^5 \theta _{\alpha \beta } m_{\alpha } \left(\eta _{\alpha \beta }+\theta _{\alpha \beta
   }\right){}^2}\\
\mathcal{A}^{2133,1}_{\i\i}=-\frac{64 k \eta _{\alpha \beta }^2 \left(\theta _{\alpha \beta }-1\right) n_{\alpha } n_{\beta } T_{\alpha } \left(4 \eta _{\alpha \beta }-5 \theta _{\alpha \beta }+9\right)}{15 \left(\eta _{\alpha \beta
   }+1\right){}^5 \theta _{\alpha \beta } m_{\alpha } \left(\eta _{\alpha \beta }+\theta _{\alpha \beta }\right)}\\
\mathcal{A}^{2143,1}_{\i\i}=-\frac{128 k \eta _{\alpha \beta }^2 \left(\theta _{\alpha \beta }-1\right){}^2 n_{\alpha } n_{\beta } T_{\alpha }}{15 \left(\eta _{\alpha \beta }+1\right){}^5 \theta _{\alpha \beta } m_{\alpha } \left(\eta
   _{\alpha \beta }+\theta _{\alpha \beta }\right)}
\end{multline}

\begin{multline}
p=2,q=2:\    \mathcal{A}^{2211,1}_{\i\i}\Omega_{\i\j}^{11}+\mathcal{A}^{2221,1}_{\i\i}\Omega_{\i\j}^{12}+\mathcal{A}^{2231,1}_{\i\i}\Omega_{\i\j}^{13}+\mathcal{A}^{2241,1}_{\i\i}\Omega_{\i\j}^{14}+\mathcal{A}^{2251,1}_{\i\i}\Omega_{\i\j}^{15}+\mathcal{A}^{2222,1}_{\i\i}\Omega_{\i\j}^{22}\\
+\mathcal{A}^{2232,1}_{\i\i}\Omega_{\i\j}^{23}+\mathcal{A}^{2242,1}_{\i\i}\Omega_{\i\j}^{24}+\mathcal{A}^{2252,1}_{\i\i}\Omega_{\i\j}^{25}+\mathcal{A}^{2233,1}_{\i\i}\Omega_{\i\j}^{33}+\mathcal{A}^{2243,1}_{\i\i}\Omega_{\i\j}^{34}+\mathcal{A}^{2253,1}_{\i\i}\Omega_{\i\j}^{35}\\
\mathcal{A}^{2211,1}_{\i\i}=-\frac{2 n_{\alpha } n_{\beta } \left(-8 \eta _{\alpha \beta }^4 \left(14 \theta _{\alpha \beta }-19\right)+168 \eta _{\alpha \beta }^3 \theta _{\alpha \beta }^2-28 \eta _{\alpha \beta }^2 \theta _{\alpha \beta
   }^2 \left(5 \theta _{\alpha \beta }-11\right)+35 \eta _{\alpha \beta } \theta _{\alpha \beta }^4+40 \eta _{\alpha \beta }^5+35 \theta _{\alpha \beta }^4\right)}{3 \left(\eta _{\alpha \beta }+1\right){}^2
   \left(\eta _{\alpha \beta }+\theta _{\alpha \beta }\right){}^4}\\
\mathcal{A}^{2221,1}_{\i\i}=\frac{16 n_{\alpha } n_{\beta }}{15 \left(\eta _{\alpha \beta }+1\right){}^4 \left(\eta _{\alpha \beta }+\theta _{\alpha
   \beta }\right){}^4} \left(-68 \eta _{\alpha \beta }^6 \left(\theta _{\alpha \beta }-1\right)+4 \eta _{\alpha \beta }^5 \left(42 \theta _{\alpha \beta }^2-97 \theta _{\alpha \beta }+76\right)\right.\\
   \left.+\eta
   _{\alpha \beta }^4 \left(-189 \theta _{\alpha \beta }^3+511 \theta _{\alpha \beta }^2-376 \theta _{\alpha \beta }+306\right)+7 \eta _{\alpha \beta }^3 \theta _{\alpha \beta } \left(10 \theta _{\alpha \beta
   }^3-59 \theta _{\alpha \beta }^2+68 \theta _{\alpha \beta }+22\right)+7 \eta _{\alpha \beta }^2 \theta _{\alpha \beta }^2 \left(15 \theta _{\alpha \beta }^2-37 \theta _{\alpha \beta }+49\right)\right.\\
   \left.+35 \eta
   _{\alpha \beta } \theta _{\alpha \beta }^3 \left(2 \theta _{\alpha \beta }+1\right)+35 \theta _{\alpha \beta }^4\right)\\
\mathcal{A}^{2231,1}_{\i\i}=-\frac{16 n_{\alpha } n_{\beta }}{105 \left(\eta _{\alpha \beta }+1\right){}^5 \left(\eta _{\alpha \beta }+\theta _{\alpha \beta }\right){}^4} \left(12 \eta _{\alpha \beta }^6 \left(31 \theta _{\alpha \beta }^2-62 \theta _{\alpha \beta }+40\right)+\eta _{\alpha \beta }^5 \left(-714 \theta _{\alpha \beta }^3+2410 \theta
   _{\alpha \beta }^2-3064 \theta _{\alpha \beta }+1800\right)\right.\\
   \left.+\eta _{\alpha \beta }^4 \left(399 \theta _{\alpha \beta }^4-2128 \theta _{\alpha \beta }^3+2916 \theta _{\alpha \beta }^2-1936 \theta _{\alpha \beta
   }+1530\right)+2 \eta _{\alpha \beta }^3 \theta _{\alpha \beta } \left(329 \theta _{\alpha \beta }^3-924 \theta _{\alpha \beta }^2+465 \theta _{\alpha \beta }+612\right)\right.\\
   \left.+2 \eta _{\alpha \beta }^2 \theta
   _{\alpha \beta }^2 \left(301 \theta _{\alpha \beta }^2-504 \theta _{\alpha \beta }+656\right)+266 \eta _{\alpha \beta } \theta _{\alpha \beta }^3 \left(\theta _{\alpha \beta }+1\right)+133 \theta _{\alpha
   \beta }^4\right)\\
\mathcal{A}^{2241,1}_{\i\i}=-\frac{64 n_{\alpha } n_{\beta }}{105 \left(\eta _{\alpha \beta }+1\right){}^5 \left(\eta _{\alpha \beta }+\theta _{\alpha \beta }\right){}^4} \left(\eta _{\alpha \beta }^5 \left(27 \theta _{\alpha \beta }^3-81 \theta _{\alpha \beta }^2+100 \theta _{\alpha \beta }-46\right)+\eta _{\alpha \beta }^4 \left(-28 \theta
   _{\alpha \beta }^4+118 \theta _{\alpha \beta }^3-183 \theta _{\alpha \beta }^2+174 \theta _{\alpha \beta }-88\right)\right.\\
   \left.+\eta _{\alpha \beta }^3 \theta _{\alpha \beta } \left(-21 \theta _{\alpha \beta }^3+12
   \theta _{\alpha \beta }^2+75 \theta _{\alpha \beta }-94\right)+\eta _{\alpha \beta }^2 \theta _{\alpha \beta }^2 \left(-35 \theta _{\alpha \beta }^2+68 \theta _{\alpha \beta }-75\right)-7 \eta _{\alpha \beta }
   \theta _{\alpha \beta }^3 \left(\theta _{\alpha \beta }+3\right)-7 \theta _{\alpha \beta }^4\right)\\
\mathcal{A}^{2251,1}_{\i\i}=-\frac{32 n_{\alpha } n_{\beta }}{105 \left(\eta _{\alpha \beta }+1\right){}^5 \left(\eta _{\alpha \beta }+\theta _{\alpha \beta }\right){}^4} \left(\eta _{\alpha \beta }^4 \left(5 \theta _{\alpha \beta }^4-20 \theta _{\alpha \beta }^3+36 \theta _{\alpha \beta }^2-32 \theta _{\alpha \beta }+12\right)+4 \eta _{\alpha
   \beta }^3 \theta _{\alpha \beta } \left(3 \theta _{\alpha \beta }^2-6 \theta _{\alpha \beta }+4\right)\right.\\
   \left.+6 \eta _{\alpha \beta }^2 \theta _{\alpha \beta }^2 \left(\theta _{\alpha \beta }^2-2 \theta _{\alpha
   \beta }+2\right)+4 \eta _{\alpha \beta } \theta _{\alpha \beta }^3+\theta _{\alpha \beta }^4\right)\\
\mathcal{A}^{2222,1}_{\i\i}=-\frac{16 \eta _{\alpha \beta } n_{\alpha } n_{\beta } \left(\eta _{\alpha \beta }^3 \left(116-84 \theta _{\alpha \beta }\right)+\eta _{\alpha \beta }^2 \left(63 \theta _{\alpha \beta }^2-154 \theta _{\alpha
   \beta }+135\right)+7 \eta _{\alpha \beta } \left(13-5 \theta _{\alpha \beta }\right) \theta _{\alpha \beta }^2+16 \eta _{\alpha \beta }^4+7 \left(9-5 \theta _{\alpha \beta }\right) \theta _{\alpha \beta
   }^2\right)}{15 \left(\eta _{\alpha \beta }+1\right){}^4 \left(\eta _{\alpha \beta }+\theta _{\alpha \beta }\right){}^3}\\
\mathcal{A}^{2232,1}_{\i\i}=-\frac{32 \eta _{\alpha \beta } n_{\alpha } n_{\beta } }{105 \left(\eta _{\alpha \beta }+1\right){}^5 \left(\eta _{\alpha
   \beta }+\theta _{\alpha \beta }\right){}^3}\left(132 \eta _{\alpha \beta }^4 \left(\theta _{\alpha \beta }-1\right)+\eta _{\alpha \beta }^3 \left(-294 \theta _{\alpha \beta }^2+796 \theta _{\alpha
   \beta }-558\right)\right.\\
   \left.+\eta _{\alpha \beta }^2 \left(168 \theta _{\alpha \beta }^3-658 \theta _{\alpha \beta }^2+853 \theta _{\alpha \beta }-531\right)+14 \eta _{\alpha \beta } \theta _{\alpha \beta } \left(14
   \theta _{\alpha \beta }^2-17 \theta _{\alpha \beta }-9\right)+7 \theta _{\alpha \beta }^2 \left(19 \theta _{\alpha \beta }-27\right)\right)\\
\mathcal{A}^{2242,1}_{\i\i}=-\frac{64 \eta _{\alpha \beta } n_{\alpha } n_{\beta }}{105 \left(\eta _{\alpha \beta }+1\right){}^5 \left(\eta _{\alpha \beta }+\theta _{\alpha \beta }\right){}^3} \left(\eta _{\alpha \beta }^3 \left(31 \theta _{\alpha \beta }^2-62 \theta _{\alpha \beta }+35\right)+\eta _{\alpha \beta }^2 \left(-35 \theta _{\alpha \beta
   }^3+122 \theta _{\alpha \beta }^2-152 \theta _{\alpha \beta }+77\right)\right.\\
   \left.-2 \eta _{\alpha \beta } \theta _{\alpha \beta } \left(7 \theta _{\alpha \beta }^2+5 \theta _{\alpha \beta }-18\right)+\left(25-21 \theta
   _{\alpha \beta }\right) \theta _{\alpha \beta }^2\right)\\
\mathcal{A}^{2252,1}_{\i\i}=-\frac{128 \eta _{\alpha \beta } \left(\theta _{\alpha \beta }-1\right) n_{\alpha } n_{\beta } \left(\eta _{\alpha \beta }^2 \left(2 \theta _{\alpha \beta }^2-4 \theta _{\alpha \beta }+3\right)+2 \eta _{\alpha
   \beta } \theta _{\alpha \beta }+\theta _{\alpha \beta }^2\right)}{105 \left(\eta _{\alpha \beta }+1\right){}^5 \left(\eta _{\alpha \beta }+\theta _{\alpha \beta }\right){}^3}\\
\mathcal{A}^{2233,1}_{\i\i}=-\frac{32 \eta _{\alpha \beta }^2 n_{\alpha } n_{\beta } \left(-8 \eta _{\alpha \beta } \left(7 \theta _{\alpha \beta }-9\right)+8 \eta _{\alpha \beta }^2+35 \theta _{\alpha \beta }^2-126 \theta _{\alpha \beta
   }+99\right)}{105 \left(\eta _{\alpha \beta }+1\right){}^5 \left(\eta _{\alpha \beta }+\theta _{\alpha \beta }\right){}^2}\\
\mathcal{A}^{2243,1}_{\i\i}=-\frac{128 \eta _{\alpha \beta }^2 \left(\theta _{\alpha \beta }-1\right) n_{\alpha } n_{\beta } \left(4 \eta _{\alpha \beta }-7 \theta _{\alpha \beta }+11\right)}{105 \left(\eta _{\alpha \beta }+1\right){}^5
   \left(\eta _{\alpha \beta }+\theta _{\alpha \beta }\right){}^2}\\
\mathcal{A}^{2253,1}_{\i\i}=-\frac{128 \eta _{\alpha \beta }^2 \left(\theta _{\alpha \beta }-1\right){}^2 n_{\alpha } n_{\beta }}{105 \left(\eta _{\alpha \beta }+1\right){}^5 \left(\eta _{\alpha \beta }+\theta _{\alpha \beta }\right){}^2}  
\end{multline}

\subsection{Values of $B^{2pq}_{\i\j}$ for $p,q\leq1$}

\begin{multline}
 p=0,q=0:\   \mathcal{A}^{0011,2}_{\i\j} \Omega_{\i\j}^{11}+\mathcal{A}^{0021,2}_{\i\j} \Omega_{\i\j}^{12}+\mathcal{A}^{0022,2}_{\i\j} \Omega_{\i\j}^{22}\\
\mathcal{A}^{0011,2}_{\i\j}= \frac{32 \eta _{\alpha \beta } \theta _{\alpha \beta } n_{\alpha } n_{\beta }}{3 \left(\eta _{\alpha \beta }+1\right) \left(\eta _{\alpha \beta }+\theta _{\alpha \beta }\right)}\\
\mathcal{A}^{0021,2}_{\i\j}=-\frac{64 \eta _{\alpha \beta }^2 \left(\theta _{\alpha \beta }-1\right) n_{\alpha } n_{\beta }}{15 \left(\eta _{\alpha \beta }+1\right){}^2 \left(\eta _{\alpha \beta }+\theta _{\alpha \beta }\right)}\\
\mathcal{A}^{0022,2}_{\i\j}=-\frac{16 \eta _{\alpha \beta } n_{\alpha } n_{\beta }}{5 \left(\eta _{\alpha \beta }+1\right){}^2}
\end{multline}

\begin{multline}
 p=0,q=1:\   \mathcal{A}^{0111,2}_{\i\j} \Omega_{\i\j}^{11}+\mathcal{A}^{0121,2}_{\i\j} \Omega_{\i\j}^{12}+\mathcal{A}^{0131,2}_{\i\j} \Omega_{\i\j}^{13}+\mathcal{A}^{0122,2}_{\i\j} \Omega_{\i\j}^{22}+\mathcal{A}^{0123,2}_{\i\j} \Omega_{\i\j}^{23}\\
 \mathcal{A}^{0111,2}_{\i\j}=-\frac{32 \eta _{\alpha \beta }^2 \theta _{\alpha \beta } m_{\beta } n_{\alpha } n_{\beta }}{3 k \left(\eta _{\alpha \beta }+1\right) T_{\beta } \left(\eta _{\alpha \beta }+\theta _{\alpha \beta }\right){}^2}\\
\mathcal{A}^{0121,2}_{\i\j}=\frac{64 \eta _{\alpha \beta }^2 m_{\beta } n_{\alpha } n_{\beta } \left(\eta _{\alpha \beta } \left(2 \theta _{\alpha \beta }-1\right)+\theta _{\alpha \beta }\right)}{15 k \left(\eta _{\alpha \beta
   }+1\right){}^2 T_{\beta } \left(\eta _{\alpha \beta }+\theta _{\alpha \beta }\right){}^2}\\
\mathcal{A}^{0131,2}_{\i\j}=-\frac{128 \eta _{\alpha \beta }^3 \left(\theta _{\alpha \beta }-1\right) m_{\beta } n_{\alpha } n_{\beta }}{105 k \left(\eta _{\alpha \beta }+1\right){}^2 T_{\beta } \left(\eta _{\alpha \beta }+\theta _{\alpha
   \beta }\right){}^2}\\
\mathcal{A}^{0122,2}_{\i\j}=\frac{16 \eta _{\alpha \beta }^2 m_{\beta } n_{\alpha } n_{\beta }}{5 k \left(\eta _{\alpha \beta }+1\right){}^2 T_{\beta } \left(\eta _{\alpha \beta }+\theta _{\alpha \beta }\right)}\\
\mathcal{A}^{0132,2}_{\i\j}=-\frac{32 \eta _{\alpha \beta }^2 m_{\beta } n_{\alpha } n_{\beta }}{35 k \left(\eta _{\alpha \beta }+1\right){}^2 T_{\beta } \left(\eta _{\alpha \beta }+\theta _{\alpha \beta }\right)}
\end{multline}

\begin{multline}
 p=1,q=0:\   \mathcal{A}^{1011,2}_{\i\j} \Omega_{\i\j}^{11}+\mathcal{A}^{1021,2}_{\i\j} \Omega_{\i\j}^{12}+\mathcal{A}^{1031,2}_{\i\j} \Omega_{\i\j}^{13}+\mathcal{A}^{1022,2}_{\i\j} \Omega_{\i\j}^{22}+\mathcal{A}^{1023,2}_{\i\j} \Omega_{\i\j}^{23}+\mathcal{A}^{1033,2}_{\i\j} \Omega_{\i\j}^{33}\\
 \mathcal{A}^{1011,2}_{\i\j}=-\frac{112 k \eta _{\alpha \beta } \theta _{\alpha \beta } n_{\alpha } n_{\beta } T_{\alpha } \left(\eta _{\alpha \beta } \left(2 \theta _{\alpha \beta }-1\right)+\theta _{\alpha \beta }\right)}{3 \left(\eta
   _{\alpha \beta }+1\right){}^2 m_{\alpha } \left(\eta _{\alpha \beta }+\theta _{\alpha \beta }\right){}^2}\\
\mathcal{A}^{1021,2}_{\i\j}=\frac{224 k \eta _{\alpha \beta } n_{\alpha } n_{\beta } T_{\alpha } \left(\eta _{\alpha \beta }^2 \left(4 \theta _{\alpha \beta }^2-7 \theta _{\alpha \beta }+4\right)+\eta _{\alpha \beta } \theta _{\alpha \beta
   } \left(\theta _{\alpha \beta }+1\right)+\theta _{\alpha \beta }^2\right)}{15 \left(\eta _{\alpha \beta }+1\right){}^3 m_{\alpha } \left(\eta _{\alpha \beta }+\theta _{\alpha \beta }\right){}^2}\\
\mathcal{A}^{1031,2}_{\i\j}=-\frac{32 k \eta _{\alpha \beta }^2 \left(\theta _{\alpha \beta }-1\right) n_{\alpha } n_{\beta } T_{\alpha } \left(\eta _{\alpha \beta }^2 \left(4 \theta _{\alpha \beta }^2-8 \theta _{\alpha \beta }+5\right)+2
   \eta _{\alpha \beta } \theta _{\alpha \beta }+\theta _{\alpha \beta }^2\right)}{15 \left(\eta _{\alpha \beta }+1\right){}^4 \theta _{\alpha \beta } m_{\alpha } \left(\eta _{\alpha \beta }+\theta _{\alpha \beta
   }\right){}^2}\\
\mathcal{A}^{1022,2}_{\i\j}=\frac{56 k \eta _{\alpha \beta } n_{\alpha } n_{\beta } T_{\alpha } \left(\eta _{\alpha \beta } \left(11 \theta _{\alpha \beta }-8\right)+3 \theta _{\alpha \beta }\right)}{15 \left(\eta _{\alpha \beta
   }+1\right){}^3 m_{\alpha } \left(\eta _{\alpha \beta }+\theta _{\alpha \beta }\right)}\\
\mathcal{A}^{1032,2}_{\i\j}=-\frac{16 k \eta _{\alpha \beta } n_{\alpha } n_{\beta } T_{\alpha } \left(\eta _{\alpha \beta }^2 \left(11 \theta _{\alpha \beta }^2-22 \theta _{\alpha \beta }+14\right)+6 \eta _{\alpha \beta } \theta _{\alpha
   \beta }+3 \theta _{\alpha \beta }^2\right)}{15 \left(\eta _{\alpha \beta }+1\right){}^4 \theta _{\alpha \beta } m_{\alpha } \left(\eta _{\alpha \beta }+\theta _{\alpha \beta }\right)}\\
\mathcal{A}^{1033,2}_{\i\j}=-\frac{32 k \eta _{\alpha \beta }^2 \left(\theta _{\alpha \beta }-1\right) n_{\alpha } n_{\beta } T_{\alpha }}{5 \left(\eta _{\alpha \beta }+1\right){}^4 \theta _{\alpha \beta } m_{\alpha }}
\end{multline}

\begin{multline}
 p=1,q=1:\   \mathcal{A}^{1111,2}_{\i\j} \Omega_{\i\j}^{11}+\mathcal{A}^{1121,2}_{\i\j} \Omega_{\i\j}^{12}+\mathcal{A}^{1131,2}_{\i\j} \Omega_{\i\j}^{13}+\mathcal{A}^{1141,2}_{\i\j} \Omega_{\i\j}^{14}+\mathcal{A}^{1122,2}_{\i\j} \Omega_{\i\j}^{22}+\mathcal{A}^{1132,2}_{\i\j} \Omega_{\i\j}^{23}+\mathcal{A}^{1142,2}_{\i\j} \Omega_{\i\j}^{24}\\
 +\mathcal{A}^{1133,2}_{\i\j} \Omega_{\i\j}^{33}+\mathcal{A}^{1143,2}_{\i\j} \Omega_{\i\j}^{34}\\
\mathcal{A}^{1111,2}_{\i\j}=\frac{16 \eta _{\alpha \beta } \theta _{\alpha \beta }^2 n_{\alpha } n_{\beta } \left(\eta _{\alpha \beta } \left(18 \theta _{\alpha \beta }-7\right)+11 \theta _{\alpha \beta }\right)}{3 \left(\eta _{\alpha \beta
   }+1\right){}^2 \left(\eta _{\alpha \beta }+\theta _{\alpha \beta }\right){}^3}\\
\mathcal{A}^{1121,2}_{\i\j}=-\frac{64 \eta _{\alpha \beta } \theta _{\alpha \beta } n_{\alpha } n_{\beta } \left(\eta _{\alpha \beta }^2 \left(27 \theta _{\alpha \beta }^2-34 \theta _{\alpha \beta }+14\right)+2 \eta _{\alpha \beta } \theta
   _{\alpha \beta } \left(10 \theta _{\alpha \beta }-3\right)+7 \theta _{\alpha \beta }^2\right)}{15 \left(\eta _{\alpha \beta }+1\right){}^3 \left(\eta _{\alpha \beta }+\theta _{\alpha \beta }\right){}^3}\\
\mathcal{A}^{1131,2}_{\i\j}=\frac{32 \eta _{\alpha \beta } n_{\alpha } n_{\beta } \left(\eta _{\alpha \beta }^3 \left(108 \theta _{\alpha \beta }^3-230 \theta _{\alpha \beta }^2+173 \theta _{\alpha \beta }-35\right)+2 \eta _{\alpha \beta
   }^2 \theta _{\alpha \beta } \left(47 \theta _{\alpha \beta }^2-57 \theta _{\alpha \beta }+34\right)+\eta _{\alpha \beta } \theta _{\alpha \beta }^2 \left(37 \theta _{\alpha \beta }+11\right)+16 \theta _{\alpha
   \beta }^3\right)}{105 \left(\eta _{\alpha \beta }+1\right){}^4 \left(\eta _{\alpha \beta }+\theta _{\alpha \beta }\right){}^3}\\
\mathcal{A}^{1141,2}_{\i\j}=-\frac{64 \eta _{\alpha \beta }^2 \left(\theta _{\alpha \beta }-1\right) n_{\alpha } n_{\beta } \left(\eta _{\alpha \beta }^2 \left(4 \theta _{\alpha \beta }^2-8 \theta _{\alpha \beta }+5\right)+2 \eta _{\alpha
   \beta } \theta _{\alpha \beta }+\theta _{\alpha \beta }^2\right)}{105 \left(\eta _{\alpha \beta }+1\right){}^4 \left(\eta _{\alpha \beta }+\theta _{\alpha \beta }\right){}^3}\\
\mathcal{A}^{1122,2}_{\i\j}=-\frac{8 \eta _{\alpha \beta } \theta _{\alpha \beta } n_{\alpha } n_{\beta } \left(\eta _{\alpha \beta } \left(99 \theta _{\alpha \beta }-56\right)+43 \theta _{\alpha \beta }\right)}{15 \left(\eta _{\alpha \beta
   }+1\right){}^3 \left(\eta _{\alpha \beta }+\theta _{\alpha \beta }\right){}^2}\\
\mathcal{A}^{1132,2}_{\i\j}=\frac{32 \eta _{\alpha \beta } n_{\alpha } n_{\beta } \left(\eta _{\alpha \beta }^2 \left(99 \theta _{\alpha \beta }^2-127 \theta _{\alpha \beta }+49\right)+\eta _{\alpha \beta } \theta _{\alpha \beta } \left(71
   \theta _{\alpha \beta }-29\right)+21 \theta _{\alpha \beta }^2\right)}{105 \left(\eta _{\alpha \beta }+1\right){}^4 \left(\eta _{\alpha \beta }+\theta _{\alpha \beta }\right){}^2}\\
\mathcal{A}^{1142,2}_{\i\j}=-\frac{32 \eta _{\alpha \beta } n_{\alpha } n_{\beta } \left(\eta _{\alpha \beta }^2 \left(11 \theta _{\alpha \beta }^2-22 \theta _{\alpha \beta }+14\right)+6 \eta _{\alpha \beta } \theta _{\alpha \beta }+3
   \theta _{\alpha \beta }^2\right)}{105 \left(\eta _{\alpha \beta }+1\right){}^4 \left(\eta _{\alpha \beta }+\theta _{\alpha \beta }\right){}^2}\\
\mathcal{A}^{1133,2}_{\i\j}=\frac{32 \eta _{\alpha \beta } n_{\alpha } n_{\beta } \left(\eta _{\alpha \beta } \left(9 \theta _{\alpha \beta }-7\right)+2 \theta _{\alpha \beta }\right)}{35 \left(\eta _{\alpha \beta }+1\right){}^4 \left(\eta
   _{\alpha \beta }+\theta _{\alpha \beta }\right)}\\
\mathcal{A}^{1143,2}_{\i\j}=-\frac{64 \eta _{\alpha \beta }^2 \left(\theta _{\alpha \beta }-1\right) n_{\alpha } n_{\beta }}{35 \left(\eta _{\alpha \beta }+1\right){}^4 \left(\eta _{\alpha \beta }+\theta _{\alpha \beta }\right)}
\end{multline}

\subsection{Values for $A^{2pq}_{\i\j}$ for $p,q\leq1$}

\begin{multline}
 p=0,q=0:\   \mathcal{A}^{0011,2}_{\i\i} \Omega_{\i\j}^{11}+\mathcal{A}^{0021,2}_{\i\i} \Omega_{\i\j}^{12}+\mathcal{A}^{0022,2}_{\i\i} \Omega_{\i\j}^{22}\\
\mathcal{A}^{0011,2}_{\i\i}=-\frac{32 \eta _{\alpha \beta } n_{\alpha } n_{\beta }}{3 \left(\eta _{\alpha \beta }+1\right) \left(\eta _{\alpha \beta }+\theta _{\alpha \beta }\right)}\\
\mathcal{A}^{0021,2}_{\i\i}=-\frac{64 \eta _{\alpha \beta } \left(\theta _{\alpha \beta }-1\right) n_{\alpha } n_{\beta }}{15 \left(\eta _{\alpha \beta }+1\right){}^2 \left(\eta _{\alpha \beta }+\theta _{\alpha \beta }\right)}\\
\mathcal{A}^{0022,2}_{\i\i}=-\frac{16 n_{\alpha } n_{\beta }}{5 \left(\eta _{\alpha \beta }+1\right){}^2}
\end{multline}

\begin{multline}
 p=0,q=1:\   \mathcal{A}^{0111,2}_{\i\i} \Omega_{\i\j}^{11}+\mathcal{A}^{0121,2}_{\i\i} \Omega_{\i\j}^{12}+\mathcal{A}^{0131,2}_{\i\i} \Omega_{\i\j}^{13}+\mathcal{A}^{0122,2}_{\i\i} \Omega_{\i\j}^{22}+\mathcal{A}^{0123,2}_{\i\i} \Omega_{\i\j}^{23}\\
\mathcal{A}^{0111,2}_{\i\i}=\frac{32 \eta _{\alpha \beta } \theta _{\alpha \beta } m_{\alpha } n_{\alpha } n_{\beta }}{3 k \left(\eta _{\alpha \beta }+1\right) T_{\alpha } \left(\eta _{\alpha \beta }+\theta _{\alpha \beta }\right){}^2}\\
\mathcal{A}^{0121,2}_{\i\i}=-\frac{64 \eta _{\alpha \beta } \theta _{\alpha \beta } m_{\alpha } n_{\alpha } n_{\beta } \left(\eta _{\alpha \beta }-\theta _{\alpha \beta }+2\right)}{15 k \left(\eta _{\alpha \beta }+1\right){}^2 T_{\alpha }
   \left(\eta _{\alpha \beta }+\theta _{\alpha \beta }\right){}^2}\\
\mathcal{A}^{0131,2}_{\i\i}=-\frac{128 \eta _{\alpha \beta } \left(\theta _{\alpha \beta }-1\right) \theta _{\alpha \beta } m_{\alpha } n_{\alpha } n_{\beta }}{105 k \left(\eta _{\alpha \beta }+1\right){}^2 T_{\alpha } \left(\eta _{\alpha
   \beta }+\theta _{\alpha \beta }\right){}^2}\\
\mathcal{A}^{0122,2}_{\i\i}=\frac{16 \theta _{\alpha \beta } m_{\alpha } n_{\alpha } n_{\beta }}{5 k \left(\eta _{\alpha \beta }+1\right){}^2 T_{\alpha } \left(\eta _{\alpha \beta }+\theta _{\alpha \beta }\right)}\\
\mathcal{A}^{0132,2}_{\i\i}=-\frac{32 \theta _{\alpha \beta } m_{\alpha } n_{\alpha } n_{\beta }}{35 k \left(\eta _{\alpha \beta }+1\right){}^2 T_{\alpha } \left(\eta _{\alpha \beta }+\theta _{\alpha \beta }\right)}
\end{multline}

\begin{multline}
 p=1,q=0:\   \mathcal{A}^{1011,2}_{\i\i} \Omega_{\i\j}^{11}+\mathcal{A}^{1021,2}_{\i\i} \Omega_{\i\j}^{12}+\mathcal{A}^{1031,2}_{\i\i} \Omega_{\i\j}^{13}+\mathcal{A}^{1022,2}_{\i\i} \Omega_{\i\j}^{22}+\mathcal{A}^{1023,2}_{\i\i} \Omega_{\i\j}^{23}+\mathcal{A}^{1033,2}_{\i\i} \Omega_{\i\j}^{33}\\
 \mathcal{A}^{1011,2}_{\i\i}=\frac{112 k \eta _{\alpha \beta } n_{\alpha } n_{\beta } T_{\alpha } \left(\eta _{\alpha \beta }^2 \left(-\left(\theta _{\alpha \beta }-1\right)\right)+\eta _{\alpha \beta } \theta _{\alpha \beta }^2+\theta
   _{\alpha \beta }^2\right)}{3 \left(\eta _{\alpha \beta }+1\right){}^2 \theta _{\alpha \beta } m_{\alpha } \left(\eta _{\alpha \beta }+\theta _{\alpha \beta }\right){}^2}\\
\mathcal{A}^{1021,2}_{\i\i}=\frac{224 k \eta _{\alpha \beta } n_{\alpha } n_{\beta } T_{\alpha } \left(\eta _{\alpha \beta }^2 \left(-3 \theta _{\alpha \beta }^2+6 \theta _{\alpha \beta }-4\right)+\eta _{\alpha \beta } \theta _{\alpha \beta
   } \left(\theta _{\alpha \beta }^2-\theta _{\alpha \beta }-2\right)+\left(\theta _{\alpha \beta }-2\right) \theta _{\alpha \beta }^2\right)}{15 \left(\eta _{\alpha \beta }+1\right){}^3 \theta _{\alpha \beta }
   m_{\alpha } \left(\eta _{\alpha \beta }+\theta _{\alpha \beta }\right){}^2}\\
\mathcal{A}^{1031,2}_{\i\i}=-\frac{32 k \eta _{\alpha \beta } \left(\theta _{\alpha \beta }-1\right) n_{\alpha } n_{\beta } T_{\alpha } \left(\eta _{\alpha \beta }^2 \left(4 \theta _{\alpha \beta }^2-8 \theta _{\alpha \beta }+5\right)+2
   \eta _{\alpha \beta } \theta _{\alpha \beta }+\theta _{\alpha \beta }^2\right)}{15 \left(\eta _{\alpha \beta }+1\right){}^4 \theta _{\alpha \beta } m_{\alpha } \left(\eta _{\alpha \beta }+\theta _{\alpha \beta
   }\right){}^2}\\
\mathcal{A}^{1022,2}_{\i\i}=-\frac{56 k n_{\alpha } n_{\beta } T_{\alpha } \left(8 \eta _{\alpha \beta }^2 \left(\theta _{\alpha \beta }-1\right)-3 \eta _{\alpha \beta } \theta _{\alpha \beta }^2-3 \theta _{\alpha \beta }^2\right)}{15
   \left(\eta _{\alpha \beta }+1\right){}^3 \theta _{\alpha \beta } m_{\alpha } \left(\eta _{\alpha \beta }+\theta _{\alpha \beta }\right)}\\
\mathcal{A}^{1032,2}_{\i\i}=-\frac{16 k n_{\alpha } n_{\beta } T_{\alpha } \left(\eta _{\alpha \beta }^2 \left(11 \theta _{\alpha \beta }^2-22 \theta _{\alpha \beta }+14\right)+6 \eta _{\alpha \beta } \theta _{\alpha \beta }+3 \theta
   _{\alpha \beta }^2\right)}{15 \left(\eta _{\alpha \beta }+1\right){}^4 \theta _{\alpha \beta } m_{\alpha } \left(\eta _{\alpha \beta }+\theta _{\alpha \beta }\right)}\\
\mathcal{A}^{1033,2}_{\i\i}=-\frac{32 k \eta _{\alpha \beta } \left(\theta _{\alpha \beta }-1\right) n_{\alpha } n_{\beta } T_{\alpha }}{5 \left(\eta _{\alpha \beta }+1\right){}^4 \theta _{\alpha \beta } m_{\alpha }}
\end{multline}

\begin{multline}
 p=1,q=1:\   \mathcal{A}^{1111,2}_{\i\i} \Omega_{\i\j}^{11}+\mathcal{A}^{1121,2}_{\i\i} \Omega_{\i\j}^{12}+\mathcal{A}^{1131,2}_{\i\i} \Omega_{\i\j}^{13}+\mathcal{A}^{1141,2}_{\i\i} \Omega_{\i\j}^{14}+\mathcal{A}^{1122,2}_{\i\i} \Omega_{\i\j}^{22}+\mathcal{A}^{1132,2}_{\i\i} \Omega_{\i\j}^{23}+\mathcal{A}^{1142,2}_{\i\i} \Omega_{\i\j}^{24}\\
 +\mathcal{A}^{1133,2}_{\i\i} \Omega_{\i\j}^{33}+\mathcal{A}^{1143,2}_{\i\i} \Omega_{\i\j}^{34}\\
\mathcal{A}^{1111,2}_{\i\i}=-\frac{16 \eta _{\alpha \beta } n_{\alpha } n_{\beta } \left(\eta _{\alpha \beta }^2 \left(11-7 \theta _{\alpha \beta }\right)+7 \eta _{\alpha \beta } \theta _{\alpha \beta }^2+4 \eta _{\alpha \beta }^3+7 \theta
   _{\alpha \beta }^2\right)}{3 \left(\eta _{\alpha \beta }+1\right){}^2 \left(\eta _{\alpha \beta }+\theta _{\alpha \beta }\right){}^3}\\
\mathcal{A}^{1121,2}_{\i\i}=-\frac{32 \eta _{\alpha \beta } n_{\alpha } n_{\beta } \left(19 \eta _{\alpha \beta }^3 \left(\theta _{\alpha \beta }-1\right)+\eta _{\alpha \beta }^2 \left(-28 \theta _{\alpha \beta }^2+61 \theta _{\alpha \beta
   }-47\right)+7 \eta _{\alpha \beta } \theta _{\alpha \beta } \left(\theta _{\alpha \beta }^2-3 \theta _{\alpha \beta }-2\right)+7 \left(\theta _{\alpha \beta }-3\right) \theta _{\alpha \beta }^2\right)}{15
   \left(\eta _{\alpha \beta }+1\right){}^3 \left(\eta _{\alpha \beta }+\theta _{\alpha \beta }\right){}^3}\\
\mathcal{A}^{1131,2}_{\i\i}=-\frac{32 \eta _{\alpha \beta } n_{\alpha } n_{\beta }}{105 \left(\eta _{\alpha \beta }+1\right){}^4 \left(\eta _{\alpha \beta }+\theta _{\alpha \beta }\right){}^3} \left(2 \eta _{\alpha \beta }^3 \left(33 \theta _{\alpha \beta }^2-66 \theta _{\alpha \beta }+41\right)+\eta _{\alpha \beta }^2 \left(-42 \theta _{\alpha
   \beta }^3+164 \theta _{\alpha \beta }^2-191 \theta _{\alpha \beta }+117\right)\right.\\
   \left.+2 \eta _{\alpha \beta } \theta _{\alpha \beta } \left(-14 \theta _{\alpha \beta }^2+15 \theta _{\alpha \beta
   }+23\right)+\left(37-21 \theta _{\alpha \beta }\right) \theta _{\alpha \beta }^2\right)\\
\mathcal{A}^{1141,2}_{\i\i}=-\frac{64 \eta _{\alpha \beta } \left(\theta _{\alpha \beta }-1\right) n_{\alpha } n_{\beta } \left(\eta _{\alpha \beta }^2 \left(4 \theta _{\alpha \beta }^2-8 \theta _{\alpha \beta }+5\right)+2 \eta _{\alpha
   \beta } \theta _{\alpha \beta }+\theta _{\alpha \beta }^2\right)}{105 \left(\eta _{\alpha \beta }+1\right){}^4 \left(\eta _{\alpha \beta }+\theta _{\alpha \beta }\right){}^3}\\
\mathcal{A}^{1122,2}_{\i\i}=-\frac{8 n_{\alpha } n_{\beta } \left(\eta _{\alpha \beta }^2 \left(78-56 \theta _{\alpha \beta }\right)+21 \eta _{\alpha \beta } \theta _{\alpha \beta }^2+22 \eta _{\alpha \beta }^3+21 \theta _{\alpha \beta
   }^2\right)}{15 \left(\eta _{\alpha \beta }+1\right){}^3 \left(\eta _{\alpha \beta }+\theta _{\alpha \beta }\right){}^2}\\
\mathcal{A}^{1132,2}_{\i\i}=-\frac{32 n_{\alpha } n_{\beta } \left(50 \eta _{\alpha \beta }^3 \left(\theta _{\alpha \beta }-1\right)+\eta _{\alpha \beta }^2 \left(-49 \theta _{\alpha \beta }^2+127 \theta _{\alpha \beta }-99\right)-21 \eta
   _{\alpha \beta } \theta _{\alpha \beta } \left(\theta _{\alpha \beta }+1\right)-21 \theta _{\alpha \beta }^2\right)}{105 \left(\eta _{\alpha \beta }+1\right){}^4 \left(\eta _{\alpha \beta }+\theta _{\alpha
   \beta }\right){}^2}\\
\mathcal{A}^{1142,2}_{\i\i}=-\frac{32 n_{\alpha } n_{\beta } \left(\eta _{\alpha \beta }^2 \left(11 \theta _{\alpha \beta }^2-22 \theta _{\alpha \beta }+14\right)+6 \eta _{\alpha \beta } \theta _{\alpha \beta }+3 \theta _{\alpha \beta
   }^2\right)}{105 \left(\eta _{\alpha \beta }+1\right){}^4 \left(\eta _{\alpha \beta }+\theta _{\alpha \beta }\right){}^2}\\
\mathcal{A}^{1133,2}_{\i\i}=-\frac{32 \eta _{\alpha \beta } n_{\alpha } n_{\beta } \left(2 \eta _{\alpha \beta }-7 \theta _{\alpha \beta }+9\right)}{35 \left(\eta _{\alpha \beta }+1\right){}^4 \left(\eta _{\alpha \beta }+\theta _{\alpha
   \beta }\right)}\\
\mathcal{A}^{1143,2}_{\i\i}=-\frac{64 \eta _{\alpha \beta } \left(\theta _{\alpha \beta }-1\right) n_{\alpha } n_{\beta }}{35 \left(\eta _{\alpha \beta }+1\right){}^4 \left(\eta _{\alpha \beta }+\theta _{\alpha \beta }\right)}
\end{multline}

\section{The 21N-moment coefficients with the Coulomb potential}
\label{sec:multi-temp_coeffs_coulomb}

Following Zhdanov\cite{zhdanov_transport_2002}, we choose the value of the Chapman-Cowling integrals $\Omega^{lr}_{\alpha\beta}$ for Coulomb potential with Debye cutoff as follows
\begin{equation}
 \Omega_{\i\j}^{lr}=l(r-1)!\left(\frac{2\pi}{\gamma_{\i\j}}\right)^{1/2}\frac{\Delta_{\i\j}^2}{4}\ln\Lambda_{\i\j}, \label{eq:zhdanov21Ncrosssections}
\end{equation}
where the Coulomb logarithm $\ln\Lambda_{\i\j}$ is given by\footnote{An earlier version of the manuscript contained a typographical erro in the value of $\Lambda_{\i\j}$. The factor should be 3 and not 12. The error is sincerely regretted. In addition, one could also use $1/\gamma^2$ instead of 3, where $\gamma$ is the Euler-Mascheroni constant. They are both permissible as they come from different approaches to calculating the mean distance of closest approach. }
\begin{equation}
 \Lambda_{\i\j}=\frac{3\lambda_D}{\Delta_{\i\j}},
\end{equation}
where the Debye length $\lambda_D$ is given by
\begin{equation}
 \lambda_D^{-2}=\sum_\alpha \frac{ n_\i Z_\i^2 e^2}{\epsilon_0 kT_\i},\nonumber
\end{equation}
and the mean distance of closest approach is given by
\begin{equation}
 \Delta_{\i\j}=\frac{|Z_\i Z_\j|e^2}{4\pi\epsilon_0}\frac{\gamma_{\i\j}}{\mu_{\i\j}}
\end{equation}
where $\gamma_{\i\j}$ and $\mu_{\i\j}$ are the reduced-$\gamma$ ($=\gamma_\i\gamma_\j/(\gamma_\i+\gamma_\j)$) and reduced mass respectively.

Although, this form of cross section is not fully accurate (there are some corrections to the Coulomb logarithm which can be found in Refs.\cite{liboff_transport_1959,capitelli_transport_2013,raghunathan_generalized_2021}), it is a reasonable approximation to the order of the Coulomb logarithm within the 21N-moment scheme, and they allow us to simplify the coefficients in the previous section significantly. 
Furthermore, we define a $\lambda_{\i\j,coulomb}$ such that 
\begin{equation}
  \lambda_{\i\j,coulomb}=\left(\frac{2\pi}{\gamma_{\i\j}}\right)^{1/2}\frac{\Delta_{\i\j}^2}{4}\ln\Lambda_{\i\j}.
\end{equation}
This value of $\lambda_{\i\j,coulomb}$ will remain a common factor in all Coulomb collision coefficients. \footnote{An earlier version of this article contained an extra factor of $\sqrt{\pi}$ in Eqs.\,(45) and (48). The error is  sincerely regretted.}
We now present the coefficients divided by $\lambda_{\i\j,coulomb}$ for clarity, as follows.

\subsection{Value of $C^{010}_{\i\j}/\lambda_{\i\j,coulomb}$ for $p,q=0$}

\begin{equation}
p=1,q=0:\   -\frac{16 k \eta _{\alpha \beta } \left(\theta _{\alpha \beta }-1\right) n_{\alpha } n_{\beta } T_{\alpha }}{\left(\eta _{\alpha \beta }+1\right){}^2 \theta _{\alpha \beta }}
\end{equation}

\subsection{Values of $B^{1pq}_{\i\j}/\lambda_{\i\j,coulomb}$ for $p,q\leq2$}

\begin{align}
 p=0,q=0:&\  \frac{16 \eta _{\alpha \beta } n_{\alpha } n_{\beta }}{3 \eta _{\alpha \beta }+3} \\
 p=0,q=1:&\ -\frac{16 \eta _{\alpha \beta }^2 m_{\beta } n_{\alpha } n_{\beta }}{5 k \left(\eta _{\alpha \beta }+1\right) T_{\beta } \left(\eta _{\alpha \beta }+\theta _{\alpha \beta }\right)} \\
 p=1,q=0:&\ -\frac{8 k \eta _{\alpha \beta } n_{\alpha } n_{\beta } T_{\alpha } \left(\eta _{\alpha \beta } \left(3 \theta _{\alpha \beta }-2\right)+\theta _{\alpha \beta }\right)}{\left(\eta _{\alpha \beta }+1\right){}^2
   m_{\alpha } \left(\eta _{\alpha \beta }+\theta _{\alpha \beta }\right)} \\
  p=1,q=1:&\ \frac{24 \eta _{\alpha \beta } \theta _{\alpha \beta } n_{\alpha } n_{\beta } \left(\eta _{\alpha \beta } \left(5 \theta _{\alpha \beta }-2\right)+3 \theta _{\alpha \beta }\right)}{5 \left(\eta _{\alpha \beta
   }+1\right){}^2 \left(\eta _{\alpha \beta }+\theta _{\alpha \beta }\right){}^2} \\
   p=0,q=2:&\ \frac{8 \eta _{\alpha \beta }^3 m_{\beta }^2 n_{\alpha } n_{\beta }}{7 k^2 \left(\eta _{\alpha \beta }+1\right) T_{\beta }^2 \left(\eta _{\alpha \beta }+\theta _{\alpha \beta }\right){}^2} \\
   p=2,q=0:&\    \frac{20 k^2 \eta _{\alpha \beta } \theta _{\alpha \beta } n_{\alpha } n_{\beta } T_{\alpha }^2 \left(\eta _{\alpha \beta } \left(5 \theta _{\alpha \beta }-4\right)+\theta _{\alpha \beta }\right)}{\left(\eta
   _{\alpha \beta }+1\right){}^2 m_{\alpha }^2 \left(\eta _{\alpha \beta }+\theta _{\alpha \beta }\right){}^2} \\
   p=1,q=2:&\   -\frac{12 \eta _{\alpha \beta }^2 \theta _{\alpha \beta } m_{\beta } n_{\alpha } n_{\beta } \left(\eta _{\alpha \beta } \left(7 \theta _{\alpha \beta }-2\right)+5 \theta _{\alpha \beta }\right)}{7 k \left(\eta
   _{\alpha \beta }+1\right){}^2 T_{\beta } \left(\eta _{\alpha \beta }+\theta _{\alpha \beta }\right){}^3} \\
   p=2,q=1:&\   -\frac{20 k \eta _{\alpha \beta } \theta _{\alpha \beta }^2 n_{\alpha } n_{\beta } T_{\alpha } \left(\eta _{\alpha \beta } \left(7 \theta _{\alpha \beta }-4\right)+3 \theta _{\alpha \beta }\right)}{\left(\eta
   _{\alpha \beta }+1\right){}^2 m_{\alpha } \left(\eta _{\alpha \beta }+\theta _{\alpha \beta }\right){}^3} \\
   p=2,q=2:&\    \frac{10 \eta _{\alpha \beta } \theta _{\alpha \beta }^3 n_{\alpha } n_{\beta } \left(\eta _{\alpha \beta } \left(9 \theta _{\alpha \beta }-4\right)+5 \theta _{\alpha \beta }\right)}{\left(\eta _{\alpha \beta
   }+1\right){}^2 \left(\eta _{\alpha \beta }+\theta _{\alpha \beta }\right){}^4} 
\end{align}

\subsection{Values of $A^{1pq}_{\i\j}/\lambda_{\i\j,coulomb}$ for $p,q\leq2$}

\begin{align}
 p=0,q=0:&\  -\frac{16 n_{\alpha } n_{\beta }}{3 \eta _{\alpha \beta }+3} \\
 p=0,q=1:&\ \frac{16 \theta _{\alpha \beta } m_{\alpha } n_{\alpha } n_{\beta }}{5 k \left(\eta _{\alpha \beta }+1\right) T_{\alpha } \left(\eta _{\alpha \beta }+\theta _{\alpha \beta }\right)} \\
 p=1,q=0:&\ -\frac{8 k n_{\alpha } n_{\beta } T_{\alpha } \left(10 \eta _{\alpha \beta }^2 \left(\theta _{\alpha \beta }-1\right)+\eta _{\alpha \beta } \left(\theta _{\alpha \beta }-4\right) \theta _{\alpha \beta }-3 \theta
   _{\alpha \beta }^2\right)}{3 \left(\eta _{\alpha \beta }+1\right){}^2 \theta _{\alpha \beta } m_{\alpha } \left(\eta _{\alpha \beta }+\theta _{\alpha \beta }\right)} \\
  p=1,q=1:&\ -\frac{8 n_{\alpha } n_{\beta } \left(\eta _{\alpha \beta }^2 \left(52-6 \theta _{\alpha \beta }\right)+\eta _{\alpha \beta } \theta _{\alpha \beta } \left(9 \theta _{\alpha \beta }+20\right)+30 \eta _{\alpha
   \beta }^3+13 \theta _{\alpha \beta }^2\right)}{15 \left(\eta _{\alpha \beta }+1\right){}^2 \left(\eta _{\alpha \beta }+\theta _{\alpha \beta }\right){}^2} \\
   p=0,q=2:&\ -\frac{8 \theta _{\alpha \beta }^2 m_{\alpha }^2 n_{\alpha } n_{\beta }}{7 k^2 \left(\eta _{\alpha \beta }+1\right) T_{\alpha }^2 \left(\eta _{\alpha \beta }+\theta _{\alpha \beta }\right){}^2} \\
   p=2,q=0:&\    \frac{4 k^2 n_{\alpha } n_{\beta } T_{\alpha }^2 \left(28 \eta _{\alpha \beta }^2 \left(\theta _{\alpha \beta }-1\right)+\eta _{\alpha \beta } \theta _{\alpha \beta } \left(3 \theta _{\alpha \beta }-8\right)-5
   \theta _{\alpha \beta }^2\right)}{\left(\eta _{\alpha \beta }+1\right){}^2 m_{\alpha }^2 \left(\eta _{\alpha \beta }+\theta _{\alpha \beta }\right){}^2} \\
   p=1,q=2:&\   \frac{4 \theta _{\alpha \beta } m_{\alpha } n_{\alpha } n_{\beta } \left(-2 \eta _{\alpha \beta }^2 \left(\theta _{\alpha \beta }-59\right)+\eta _{\alpha \beta } \theta _{\alpha \beta } \left(19 \theta _{\alpha
   \beta }+36\right)+84 \eta _{\alpha \beta }^3+23 \theta _{\alpha \beta }^2\right)}{35 k \left(\eta _{\alpha \beta }+1\right){}^2 T_{\alpha } \left(\eta _{\alpha \beta }+\theta _{\alpha \beta }\right){}^3} \\
   p=2,q=1:&\   -\frac{4 k n_{\alpha } n_{\beta } T_{\alpha }}{15 \left(\eta _{\alpha \beta }+1\right){}^2 \theta _{\alpha \beta } m_{\alpha } \left(\eta _{\alpha \beta }+\theta _{\alpha \beta }\right){}^3} \left(280 \eta _{\alpha \beta }^4 \left(\theta _{\alpha \beta }-1\right)+84 \eta _{\alpha \beta }^3 \left(\theta _{\alpha \beta }-4\right) \theta _{\alpha \beta }\right.\nonumber\\
   &\left.+348
   \eta _{\alpha \beta }^2 \left(\theta _{\alpha \beta }-2\right) \theta _{\alpha \beta }^2+\eta _{\alpha \beta } \theta _{\alpha \beta }^3 \left(19 \theta _{\alpha \beta }-184\right)-69 \theta _{\alpha \beta
   }^4\right) \\
   p=2,q=2:&\    -\frac{2 n_{\alpha } n_{\beta }}{105 \left(\eta _{\alpha \beta }+1\right){}^2 \left(\eta _{\alpha \beta }+\theta _{\alpha \beta }\right){}^4} \left(-56 \eta _{\alpha \beta }^4 \left(2 \theta _{\alpha \beta }-59\right)+8 \eta _{\alpha \beta }^3 \theta _{\alpha \beta } \left(303 \theta _{\alpha \beta }+380\right)\right.\nonumber\\
   &\left.+4 \eta
   _{\alpha \beta }^2 \left(1329-139 \theta _{\alpha \beta }\right) \theta _{\alpha \beta }^2+\eta _{\alpha \beta } \theta _{\alpha \beta }^3 \left(233 \theta _{\alpha \beta }+1288\right)+1400 \eta _{\alpha \beta
   }^5+433 \theta _{\alpha \beta }^4\right) 
\end{align}

\subsection{Values of $B^{2pq}_{\i\j}/\lambda_{\i\j,coulomb}$ for $p,q\leq1$}

\begin{align}
 p=0,q=0:&\  \frac{32 \eta _{\alpha \beta } n_{\alpha } n_{\beta } \left(\eta _{\alpha \beta } \left(3 \theta _{\alpha \beta }-1\right)+2 \theta _{\alpha \beta }\right)}{15 \left(\eta _{\alpha \beta }+1\right){}^2 \left(\eta
   _{\alpha \beta }+\theta _{\alpha \beta }\right)} \\
 p=0,q=1:&\ -\frac{32 \eta _{\alpha \beta }^2 m_{\beta } n_{\alpha } n_{\beta } \left(\eta _{\alpha \beta } \left(5 \theta _{\alpha \beta }-1\right)+4 \theta _{\alpha \beta }\right)}{35 k \left(\eta _{\alpha \beta
   }+1\right){}^2 T_{\beta } \left(\eta _{\alpha \beta }+\theta _{\alpha \beta }\right){}^2} \\
 p=1,q=0:&\ -\frac{32 k \eta _{\alpha \beta } \theta _{\alpha \beta } n_{\alpha } n_{\beta } T_{\alpha } \left(\eta _{\alpha \beta } \left(5 \theta _{\alpha \beta }-3\right)+2 \theta _{\alpha \beta }\right)}{5 \left(\eta
   _{\alpha \beta }+1\right){}^2 m_{\alpha } \left(\eta _{\alpha \beta }+\theta _{\alpha \beta }\right){}^2} \\
  p=1,q=1:&\ \frac{32 \eta _{\alpha \beta } \theta _{\alpha \beta }^2 n_{\alpha } n_{\beta } \left(\eta _{\alpha \beta } \left(7 \theta _{\alpha \beta }-3\right)+4 \theta _{\alpha \beta }\right)}{7 \left(\eta _{\alpha \beta
   }+1\right){}^2 \left(\eta _{\alpha \beta }+\theta _{\alpha \beta }\right){}^3} 
\end{align}

\subsection{Values for $A^{2pq}_{\i\j}/\lambda_{\i\j,coulomb}$ for $p,q\leq1$}

\begin{align}
 p=0,q=0:&\  -\frac{32 n_{\alpha } n_{\beta } \left(2 \eta _{\alpha \beta } \left(\theta _{\alpha \beta }+3\right)+5 \eta _{\alpha \beta }^2+3 \theta _{\alpha \beta }\right)}{15 \left(\eta _{\alpha \beta }+1\right){}^2
   \left(\eta _{\alpha \beta }+\theta _{\alpha \beta }\right)} \\
 p=0,q=1:&\ \frac{32 \theta _{\alpha \beta } m_{\alpha } n_{\alpha } n_{\beta } \left(2 \eta _{\alpha \beta } \left(\theta _{\alpha \beta }+4\right)+7 \eta _{\alpha \beta }^2+3 \theta _{\alpha \beta }\right)}{35 k \left(\eta
   _{\alpha \beta }+1\right){}^2 T_{\alpha } \left(\eta _{\alpha \beta }+\theta _{\alpha \beta }\right){}^2} \\
 p=1,q=0:&\ -\frac{16 k n_{\alpha } n_{\beta } T_{\alpha } \left(35 \eta _{\alpha \beta }^3 \left(\theta _{\alpha \beta }-1\right)+7 \eta _{\alpha \beta }^2 \left(\theta _{\alpha \beta }-4\right) \theta _{\alpha \beta }+2
   \eta _{\alpha \beta } \left(\theta _{\alpha \beta }-16\right) \theta _{\alpha \beta }^2-9 \theta _{\alpha \beta }^3\right)}{15 \left(\eta _{\alpha \beta }+1\right){}^2 \theta _{\alpha \beta } m_{\alpha }
   \left(\eta _{\alpha \beta }+\theta _{\alpha \beta }\right){}^2} \\
  p=1,q=1:&\ -\frac{16 n_{\alpha } n_{\beta } \left(21 \eta _{\alpha \beta }^3 \left(\theta _{\alpha \beta }+13\right)+3 \eta _{\alpha \beta }^2 \theta _{\alpha \beta } \left(39 \theta _{\alpha \beta }+74\right)+2 \eta
   _{\alpha \beta } \theta _{\alpha \beta }^2 \left(13 \theta _{\alpha \beta }+105\right)+140 \eta _{\alpha \beta }^4+51 \theta _{\alpha \beta }^3\right)}{105 \left(\eta _{\alpha \beta }+1\right){}^2 \left(\eta
   _{\alpha \beta }+\theta _{\alpha \beta }\right){}^3} 
\end{align}

%

\section{Summary}
\label{sec:summary}

In this article, we have provided an overview of the calculation of the collisional coefficients, and then provided the collisional coefficients for the Boltzmann collision operator for the $21N$-moment multi-temperature case, both generally and specifically for the Coulomb potential with Debye cutoff. This fills an important gap in existing literature. Furthermore, we hope that these coefficients will lead to an easy implementation in existing state-of-the-art plasma fluid packages.

\bibliography{bibliography}{}
\bibliographystyle{vancouver}

\end{document}